\documentclass[referee,useAMS,usenatbib]{mn2e}%

\usepackage{amsmath,amssymb}

\usepackage[mathscr]{euscript} 

\usepackage{subeqn, wasysym}
\usepackage{natbib}
\usepackage{graphicx}
\usepackage{subcaption}
\usepackage{epstopdf}
\usepackage{placeins}
\usepackage{float}
\usepackage{comment}
\usepackage{xcolor}
\usepackage{soul}

\newcommand{\beq}{\begin{equation}}
\newcommand{\eeq}{\end{equation}}

\newcommand{\bfr}{\mbox{\boldmath $r$}}

\newcommand{\bfu}{\mbox{\boldmath $u$}}

\newcommand{\cross}{\mbox{\boldmath $\times$}}

\newcommand{\p}{\mbox{$\partial$}}
\newcommand{\tk}{\mbox{$T_{\rm kep}$}}
\newcommand{\ts}{\mbox{$T_{\rm sec}$}}

\newcommand{\scrr}{{\cal R}}

\newcommand{\rmd}{{\rm d}}

\newcommand{\mdot}{\mbox{$M_\bullet$} }
\newcommand{\msun}{\mbox{$M_{\odot} $}}

\newcommand{\rc}{\mbox{$r_{\rm c} $}}
\newcommand{\Mc}{\mbox{$ M_{\rm c} $}}

\begin{document}

\title[Cusp deformation due to a gas disc]{Deformation of the Galactic Centre stellar cusp due to the gravity of a growing gas disc}
\author[Kaur and Sridhar]{Karamveer Kaur$^1$ and S. Sridhar$^2$\\
Raman Research Institute, Sadashivanagar, Bangalore 560 080, India\\
$^{1}$~karamveer@rri.res.in $\quad^{2}$~ssridhar@rri.res.in\\}
\maketitle

\begin{abstract}
The nuclear star cluster surrounding the massive black hole at the Galactic Centre consists of young and old stars, with most of the stellar mass in an extended, cuspy distribution of old stars. The compact cluster of young stars was probably born in situ in a massive accretion disc around the black hole. We investigate the effect of the growing gravity of the disc on the orbits of the old stars, using an integrable model of the deformation of a spherical star cluster with anisotropic velocity dispersions. A formula for the perturbed phase space distribution function is derived using linear theory, and new density and surface density profiles are computed. The cusp undergoes a spheroidal deformation with the flattening increasing strongly at smaller distances from the black hole; the intrinsic axis ratio $\sim 0.8$ at $\sim 0.15~\mbox{pc}$. Stellar orbits are deformed such that they spend more
time near the disc plane and sample the dense inner parts of the disc; 
this could result in enhanced stripping of the envelopes of red giant stars. Linear theory accounts only for orbits whose apsides circulate. The non-linear theory of adiabatic capture into resonance is needed to understand orbits whose apsides librate. The mechanism is a generic dynamical process, and it may be common in galactic nuclei.
 \end{abstract}


\section{Introduction}

There is strong evidence that the Galactic centre (GC) source Sgr~A$^*$ 
is a massive black hole (MBH) with mass of about $4\times 10^6~\msun$, 
embedded in a nuclear star cluster (NSC) of $2.5\times 10^7~\msun$ with a half-light radius of about $4~\mbox{pc}$, consisting of both late-type (old, $> 1$~Gyr) and early-type (young, $< 10$~Myr) stars \citep{geg10,sfk14,bgs16,gpe17}. The first high angular resolution observations seemed to imply that the old stars were distributed in a density cusp \citep{gso03,sea07}. But when the contamination of light from the young stars was accounted for, the old giant population appeared to have a core-like, rather than a cuspy, surface density profile \citep{bse09,dgm09,bmt10,fcg16}. Recent work  has refined our knowledge of the distribution of the old stars \citep{gsd17,sgd17}. Within about $3~\mbox{pc}$ of the MBH the density profile of resolved faint stars and sub-giants and dwarfs (inferred from diffuse light) is cuspy, and well-described by a single power-law. But red clump and brighter giant stars have a similar cuspy profile only beyond a projected radius of about $0.3~\mbox{pc}$, inside which they display a core-like surface density profile. 

There are about $200$ young stars in a compact cluster of size $\lesssim 0.5~\mbox{pc}$ around the MBH, including WR stars, O, B type main sequence stars, giants and supergiants \citep{ahh90,kgd91,gbd03,pgm06,bmt10,dlg13}. Stellar
orbits have a range of eccentricities, inclinations and orientations, with about $20\%$ in a clockwise disc that extends between about $0.03 - 0.13~\mbox{pc}$, with mean eccentricity $\sim 0.3$ \citep{ygl14}.  It has been 
suggested that all the young stars could have been born in situ in a 
starburst event in a massive, fragmenting accretion disc around the MBH \citep{lb03}. If this is the case then the young star cluster has evolved dynamically since its birth in a dense and thin accretion disc. Repeated passage of the red clump and brighter giant stars through the dense inner parts of the accretion disc could have robbed them of their envelopes, rendering the innermost stars invisible; this would explain the difference between the core-like profiles of the old giants and the cuspy profiles of old stars lacking extended envelopes \citep{ac14}. In contrast the accretion disc's gravitational field will deflect the orbits of all old stars in the same manner. What is the gravitational response of an old stellar cusp to the accumulation of gas in an accretion disc around the MBH? 
 
In this paper we address this question by constructing a simple model
of the process within the radius of influence of the MBH, $r_{\rm infl} 
\simeq 2\,\mbox{pc}$. The problem is stated in \S~2 for a non-rotating, spherical stellar cusp with anisotropic velocity dispersions, which experiences gravitational perturbations due to a growing gas disc; we argue
that disc growth is slow compared to typical apse precession periods
of cusp orbits. In \S~3 we cast the dynamical problem in terms of the secular theory of \citet{st16}, which is its natural setting. In \S~4 we derive a formula for the linear perturbation to the phase space distribution function (DF): the magnitude of the perturbation is largest for orbits that are highly inclined with respect to the disc plane; it is positive when the angle between the lines of apsides and nodes is less than $45^\circ$ and negative otherwise. This is explained in terms of the secular, adiabatic dynamics of individual orbits in the combined gravitational potentials of the cusp and disc. Linear theory accounts only for orbits whose apsides circulate. The non-linear theory of adiabatic capture into resonance is needed to understand orbits whose apsides librate. In \S~5 we use the formula for the DF to compute the oblate spheroidal deformation of the three dimensional density profile of the cusp, as well as the surface density profiles for different viewing angles. We conclude in \S~6 with a discussion of linear stability, extensions to rotating and axisymmetric cusps, and that the process studied in this paper may be common in galactic nuclei.

\section{Statement of the problem}

We are interested in describing stellar dynamics within $1~\mbox{pc}$ of a MBH of mass $\mdot = 4\times 10^6~\msun$. Let $\bfr$ and $\bfu$ be the position vector and velocity of a star, relative to the MBH. Since this region is well inside $r_{\rm infl} \simeq 2\,\mbox{pc}$, the dominant gravitational force on a star is the Newtonian $1/r^2$ attraction of the MBH.  
Hence the shortest time scale associated with a stellar orbit of semi-major
axis $a$ is its Kepler orbital period, $\tk(a) \simeq 4.7\times 10^4\,
a_{\rm pc}^{3/2}~\mbox{yr}$ where $a_{\rm pc} = \left(a/1~\mbox{pc}\right)$.

\subsection{The unperturbed stellar cusp} This is assumed to be spherically symmetric about the MBH, with a density profile
\beq
\rho_{\rm c}(r) \;=\; \frac{(3-\gamma)M_{\rm c}}{4\pi\rc^3}\,\left(\frac{\rc}{r_{}}\right)^{\gamma}\,.
\label{cusp-den}
\eeq
For the GC cusp $\gamma = 1.23 \pm 0.05\,$, and $M_{\rm c} = 10^6~\msun$ is the stellar mass within a radius $\rc = 1$ pc of the MBH \citep{gsd17,sgd17}.
The gravitational potential due to the cusp ($\gamma \neq 2$) is
\beq
\varphi_{\rm c}(r) \;=\; \frac{GM_{\rm c}}{(2-\gamma)\rc}\,\left( \frac{r}{\rc} \right)^{2-\gamma}\,,
\label{cusp-pot}
\eeq
where a constant additive term has been dropped. The cusp's spherically
symmetric gravitational field will make the apsides of Kepler orbits 
precess in a retrograde sense in their respective orbital planes. The typical apse precession period is $T_{\rm pr}^{\rm c}(a) \sim \left(\mdot/M_{\rm ca}\right)\tk(a)$, where $M_{\rm ca} = M_{\rm c}\,a_{\rm pc}^{(3 -\gamma)}$ is the mass in cusp stars inside a sphere of radius $a$. Then $T_{\rm pr}^{\rm c}(a) \sim 1.8\times 10^5\,a_{\rm pc}^{(\gamma - 3/2)}~\mbox{yr}$.  Within a parsec the apse precession period is always longer than the Kepler orbital period. We assume that the distribution of these precessing orbits is such that, at every point in space, the mean velocity vanishes but the velocity distribution is anisotropic. This anisotropy is characterized by the parameter $\beta(r) = 1 - \left(\sigma_\theta^2 + \sigma_\phi^2\right)/2\sigma_r^2$, where the $\sigma$'s are velocity dispersions along the three principal directions of a polar coordinate system centred on the MBH. When $\beta(r)$ is negative(positive) the velocity distribution is tangentially(radially) biased.

The cusp is described by a probability distribution function, $f_{\rm c}(\bfr,\bfu)$, in the six dimensional phase space, $\{\bfr, \bfu\}$. For a non-rotating system with anisotropic velocity dispersion,  Jeans theorem implies that the unperturbed DF is a function of the energy per unit mass, 
$E = u^2/2 - G\mdot/r + \varphi_{\rm c}(r)\,$, and magnitude of the angular momentum per unit mass $L = \vert \bfr\cross\bfu\vert\,$ \citep{bt08}. Let us consider the double power-law DF,
\beq
f_{\rm c}(\bfr,\bfu) \;=\; 
\begin{cases}
\displaystyle{\;\frac{A}{2\pi}\, (-E)^m \, L^n}\,, \qquad E \;<\; 0\\
\qquad\quad 0\,, \qquad\qquad\;\; E \;>\; 0\,,
\end{cases}
\label{cusp-df}
\eeq
which is composed entirely of bound orbits; $m >0$ for the DF to be 
continuous at $E = 0$. For $r \leq 1~\mbox{pc}$ the Kepler potential of the MBH dominates the cluster potential, so $E \simeq E_{\rm k} = u^2/2 - G\mdot/r =\mbox{Kepler energy}$ is a good approximation. 
Henceforth we will consider the DF of 
equation~(\ref{cusp-df}) to be a function of $E_{\rm k}$ and $L$. The reason we begin with a two--integral (anisotropic) DF, $f_c = F(E_{k}, L)$, rather than an isotropic DF, $F(E_{\rm k})$, is the following. We have to deal with the response of a Keplerian stellar system over time scales that are much longer than Kepler orbital periods. As explained in more detail in \S~3 the Kepler energy, $E_{\rm k}$, is a secular invariant for processes that vary on (secular) times scales of the order of the apse precession periods, or longer. So a DF of the form, $F(E_{\rm k})$, would remain unchanged when perturbed by secularly varying gravitational potentials. Therefore we need to begin with at least a two--integral DF, in order to study non--trivial secular response. 
 
There is one relation among the three parameters ($A,m,n$) due to the normalization of the DF, $\int f_{\rm c}\,\rmd\bfr\,\rmd\bfu = 1\,$. The density is obtained by integrating the DF over velocity space: $\rho_{\rm c}(r) = M_{\rm c}\int f_{\rm c}\,\rmd\bfu\,$, which is straightforward to do 
in the standard manner \citep{bt08}. Comparing with equation~(\ref{cusp-den}) gives two more relations between $(A,m,n)$ and $\left(r_{\rm c},\gamma\right)$. It is convenient to choose the independent parameters as $\left(r_{\rm c}, m, n\right)$ and write: 
\begin{align}
A &\;=\; \frac{3-\gamma}{4\pi\,2^{\frac{n+1}{2}}\,B_{\left(\frac{n}{2} + 1,\frac{1}{2}\right)}\,B_{\left(m+1,\frac{n+3}{2}\right)}\,\rc^{3-\gamma}\,(G\mdot)^{\gamma+n}}\,,\\[1ex]\nonumber
\gamma &\;=\; \frac{2m-n+3}{2}\,,
\label{cusp-par}
\end{align}
where $B_{(p,q)}$ is the Beta function. It is also straightforward to calculate the velocity anisotropy, $\beta = -n/2$, which is now constant.
We note that for the density to be finite, $n>-2$ (or $\beta < 1$), which 
puts an upper limit on how radially biased the double power-law DF 
of equation~(\ref{cusp-df}) can be.   

\subsection{The perturbing gas disc} 

\citet{lb03} proposed that the young stars at the GC were formed in situ, in a massive accretion disc around the MBH. As gas accumulated in the accretion disc it became gravitationally unstable in efficiently cooling regions with Toomre $Q \lesssim 1$, and fragmented into massive stars  \citep{nay06,lev07}. A thin gas disc that is supported by external irradiation prior to fragmentation can have a steep surface density, $\Sigma_{\rm d}(R) \propto R^{-3/2}$ according to \citet{lev07}. This is consistent with the steep surface density profile of the clockwise disc of young stars that lies within about $0.13~\mbox{pc}$ of the MBH 
\citep{pgm06,lgh09,bmf09,ygl14}. We assume that the mass of the progenitor gas disc grew in time from some small value to a maximum value, just before the birth of the young stars. We need to choose a mass model representing an axisymmetric, thin accretion disc with surface density profile, $\Sigma_{\rm d}(R) \propto R^{-3/2}$. The gravitational potential of this mass model should be of a simple form, to enable explicit computation of the secular perturbation it exerts on the orbits of the old cusp stars. We found the following two-component model to be a suitable three dimensional density distribution:
\beq
\rho_{\rm d}(r,\theta, t) \;=\; \frac{2}{11\, \pi} \frac{M_{\rm d}(t)}{r_{\rm d}^3}  \left( \frac{r_{\rm d}}{r_{}} \right)^{5/2} \left[\, \delta\!\left( \theta - \frac{\pi}{2} \right) \;+\; \frac{9}{16} (1- \left| \cos{\theta}  \right|)^2 \, \right]\,, 
\label{disc-den}
\eeq
where $M_{\rm d}(t)$ is the mass inside a sphere of radius $r_{\rm d} = 1~\mbox{pc}$ at time $t$. The disc consists of two components: within a sphere of radius $r$, about $73\%$ of its mass is in a razor-thin component confined to the equatorial plane; about $27\%$ is in an extended but flattened corona. It is straightforward to verify that the gravitational potential due to $\rho_{\rm d}(r,\theta,t)$ is:   
\beq
\varphi_{\rm d}(r,\theta, t) \;=\; -\frac{8}{11} \frac{G \, M_{\rm d}(t)}{r_{\rm d}} \left(\frac{r_{\rm d}}{r_{}}\right)^{1/2} \left[\,\frac{9 \left(33 \,+\, \cos^2{\theta}\right)}{100} \;-\;  \frac{| \cos{\theta} |}{2}\,\right]\,.
\label{disc-pot}
\eeq

We are interested in determining the perturbation caused by the time-dependent disc potential of equation~(\ref{disc-pot}) to the DF of equation~(\ref{cusp-df}). In order to do this we assume that $M_{\rm d}(t)$
grows monotonically on a time scale, $T_{\rm grow}$, to its maximum value, $M_{\rm dm}$, just before the birth of the young stars. We now estimate 
$M_{\rm dm}$ and $T_{\rm grow}$:
\begin{itemize}

\item[]\emph{Disc mass:} A circumnuclear disc (CND), composed of molecular clouds, orbits the MBH at distances $\sim 1.5-5~\mbox{pc}$ 
\citep{gjh86,ggw87,ysb01}. The CND is presumably a remnant of the outer parts of the gas disc. If we assume that the total mass --- but not the necessarily its distribution --- in the annulus has not changed much over the last Myr, then we can estimate $M_{\rm dm}$ as follows. Since $\Sigma_{\rm d}(R) \propto R^{-3/2}$, the gas mass within $R$ is $\propto R^{1/2}$, so we set $M_{\rm dm}\left(\sqrt{5} - \sqrt{1.5}\,\right) = M_{\rm CND}\,$. Estimates of $M_{\rm CND}$ range from $10^4~\msun\,$ \citep{est11,rgw12} to $10^6~\msun\,$ \citep{css05}. Adopting a mid-value, $M_{\rm CND} \sim 10^5~\msun\,$, we infer that $M_{\rm dm} \sim 10^5~\msun\,$, which is similar to the value
suggested by \citet{nc05}.

\item[]\emph{Growth time:} $\,T_{\rm grow}$ depends on the agency that removes angular momentum from the gas flow at a radius of about a parsec. If it is accretion disc `$\alpha$-viscosity' then $T_{\rm grow}\sim \tk(1~\mbox{pc})/(\alpha\xi^2)$, where $\alpha\sim 0.3$ for gravitationally induced turbulence \citep{gam01} and $\xi \lesssim 0.1$ is the half-opening-angle of the thin disc; this gives $T_{\rm grow}\gtrsim 1.5 \times 10^7~\mbox{yr}$. If angular momentum is lost through non-axisymmetric gravitational perturbations then $T_{\rm grow}\sim \tk(1~\mbox{pc})/\delta_\varphi$ is the flow time scale, where $\delta_\varphi$ is the fractional non-axisymmetry in the gravitational potential at a radius of a parsec. Even for the pronounced $m=1$ asymmetry of the nuclear disc of M31, $\delta_\varphi \sim 10^{-3} - 10^{-2}$ \citep{cmc07}. Hence we expect, in either case, that $T_{\rm grow} \gtrsim 10^7~\mbox{yr}$ for the GC accretion disc. 

\end{itemize}

\subsection{Adiabatic nature of the perturbation}

\begin{figure}
\centering
\includegraphics[width=0.7\textwidth]{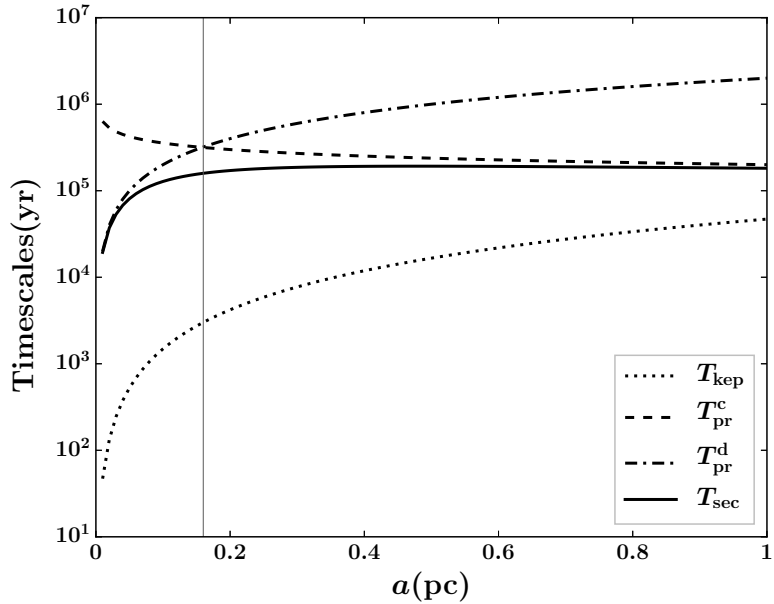}
\caption{\emph{Time scales in the problem, as functions of the semi-major axis}: The thin vertical line corresponds to $a = 0.16$ pc for which $ T_{\rm pr}^{\rm c} = T_{\rm pr}^{\rm d}$.}
\label{times}
\end{figure} 

The perturbation due to the disc contributes to both apsidal and nodal
precession. We can estimate the perturbation by imagining gas of total mass, $M_{\rm dm} = 10^5~\msun\,$, to be distributed spherically symmetric with density profile $\propto r^{-5/2}\,$, instead of being highly flattened as given by equation~(\ref{disc-den}). Such a spherically symmetric approximation to the perturbation does not cause nodal precession but contributes to retrograde apse precession over times, $T_{\rm pr}^{\rm d}(a) \sim \left(\mdot/M_{\rm da}\right)\tk(a)$, where $M_{\rm da} = 10^5\,a_{\rm pc}^{1/2}~\msun$ is the disc mass inside a sphere of radius $a$. Then $T_{\rm pr}^{\rm d}(a) \sim 2 \times 10^6\,a_{\rm pc}~\mbox{yr}$ is an increasing function of $a$. This should be compared with the retrograde apse precession period due to the cusp stars, $T_{\rm pr}^{\rm c}(a) \sim 2\times 10^5\,a_{\rm pc}^{-1/4}~\mbox{yr}$ (for a fiducial value of $\gamma = 5/4$), which is a decreasing function of $a$.
Since the apse precession due to gas and stars are
both retrograde, the net precession frequency is the sum of the individual frequencies. The corresponding  precession period then provides the 
natural time scale for secular dynamics, $T_{\rm sec}(a) = T_{\rm pr}^{\rm c}(a)\,T_{\rm pr}^{\rm d}(a)/\left[T_{\rm pr}^{\rm c}(a) + T_{\rm pr}^{\rm d}(a)\right]$. These different time scales, together with the short Kepler orbital period, $\tk(a)$, are plotted in Figure~\ref{times}. As can be seen, the net precession period, $T_{\rm sec}(a)$, is dominated by the disc mass for $a < 0.16~\mbox{pc}$ and by the cusp mass for $a > 0.16~\mbox{pc}$. This precession period attains its maximum value of about $2\times 10^5~\mbox{yr}$ within 1 pc, which is much shorter than our earlier estimate of $T_{\rm grow} \gtrsim 10^7~\mbox{yr}$, the growth time of the disc. Hence the perturbation may be assumed to be adiabatic.\footnote{Our estimates of apse precession periods accounted only for the sizes of stellar orbits (i.e. semi-major axes $a$), but not for orbital  eccentricities. Highly eccentric orbits precess very slowly --- see equation~(\ref{omega_c}) --- and the adiabatic approximation is not valid for these; this is discussed in \S~4.2.}

\section{Secular collisionless dynamics}

We have three well-separated time scales in the problem. These are the short Kepler orbital period, $\tk(a) \simeq 4.7\times 10^4\,
a_{\rm pc}^{3/2}~\mbox{yr}$; the long time scale of disc growth, 
$T_{\rm grow} \gtrsim 10^7~\mbox{yr}$; and the intermediate secular time
scale, $T_{\rm sec}(a) \lesssim 2\times 10^5~\mbox{yr}$: we always have 
$\tk(a) \ll  T_{\rm sec}(a) \ll T_{\rm grow}$ for $a \leq 1~\mbox{pc}\,$. 
In  order to study the evolution of the cusp DF over times greater than
$T_{\rm sec}(a)$, we can average the orbit of every star over the rapidly varying  Kepler orbital phase. The appropriate framework to do this is the secular theory of collisionless evolution \citep{st16}, which is briefly described below.

\subsection{General formulation of secular dynamics}

Let the stellar system be described by a normalized DF, $f(\bfr, \bfu, t)$, which satisfies the collisionless Boltzmann equation (CBE). The dynamics is governed by the Hamiltonian $H_{\rm org}(\bfr, \bfu, t)$, given in equation~(6) of \citet{st16}. Since secular dynamics corresponds to a perturbed Kepler problem, it is convenient to switch from $\{\bfr, \bfu\}$ phase space variables to the Delaunay action-angle variables, $\{I, L, L_z; w, g, h\}$. The three actions are related to the natural variables, $a = \mbox{semi-major axis}$, $e = \mbox{eccentricity}$ and $i = \mbox{inclination}$, as follows: $I\,=\,\sqrt{G\mdot a\,}\,$; $L\,=\, I\sqrt{1-e^2\,}$ the magnitude of the angular momentum; and $L_z\,=\, L\cos{i}\,$ the $z$--component of the angular momentum. The three angles conjugate to them are, respectively: $w$ the Kepler orbital phase (or mean anomaly); $g$ the angle to the periapse from the ascending node; and $h$ the longitude of the ascending node. Since the Kepler orbital energy $E_{\rm k}(I) = -1/2(G \mdot/I)^2$ depends only on the action $I$,  all the Delaunay variables except $w$ are constant in time for the unperturbed Kepler problem; $w$ itself advances at the (fast) rate $\,2\pi/\tk(a) = (GM_\bullet/a^3)^{1/2}$. Self gravity is a small perturbation to $E_{\rm k}(I)$ and so is, often, the potential due to external sources (such as the disc potential in our problem). In this case the total perturbation causes slow, secular orbital evolution and hence a natural measure of time is the `slow' time variable $\tau = \left(M_{\rm c}/ \mdot\right) t\,$. This slow dynamics is described by averaging over $w$, and its salient features are as follows:

\begin{itemize}
\item {The Hamiltonian $H_{\rm org}(\bfr, \bfu, t)$ is orbit averaged over $w$ to give the secular hamiltonian $H(I, L, L_z, g, h, \tau) = (\mdot/M_{\rm c})\oint H_{\rm org}(\bfr, \bfu, t)\,{\rmd w}/{2\pi}$ which governs the secular dynamics of the system. The phase space of the system reduces from six to five dimensions, with (`Gaussian-ring') coordinates denoted by $\scrr = \{I, L, L_z, g, h\}$. The stellar system is described by the orbit-averaged DF, $F(\scrr, \tau) = \int f(\bfr, \bfu, t)\,{\rmd w}$. Since $\int F(\scrr, \tau)\,{\rm d}\scrr = 1$, we may regard $F$ as a probability distribution function in $\scrr$--space.  }

\item {Ring orbits are governed by the secular Hamiltonian, 
$H(\scrr, \tau) = \Phi(\scrr, \tau) + \Phi^{\rm tid}(\scrr, \tau)$, which
is the sum of (scaled) contributions from self-gravity, $\Phi$, and the tidal field of external sources, $\Phi^{\rm tid}$ (relativistic effects,
included in \citet{st16} have been ignored here). The self-gravitational 
potential is related to the DF by,} 
\begin{subequations}
\begin{align}
\Phi(\scrr, \tau) &\;=\; \int F(\scrr', \tau)\,\Psi(\scrr, \scrr')
{\rm d}\scrr'\,,\qquad\quad\mbox{Ring mean--field potential};
\label{phislow-r}\\[1ex]
\Psi(\scrr, \scrr') &\;=\; -GM_\bullet\oint\oint\frac{{\rm d}w}{2\pi}\,
\frac{{\rm d}w'}{2\pi}\,\frac{1}{\left|\bfr - \bfr'\right|}\,,
\qquad\quad\mbox{`bare' inter--ring potential}. 
\label{bare-ring}
\end{align}
\end{subequations}

\item {Since the Hamiltonian is independent of $w$, its canonically conjugate action, $I$, is an integral of motion even when the Hamiltonian is time-dependent. Hence the orbit of each star is confined to the four dimensional $I = \mbox{constant}$ surface, which is equivalent to the secular conservation of its semi--major axis. On this surface the motion of each ring is governed by the following Hamiltonian equations:}
\begin{align}
\frac{{\rm d}L}{{\rm d}\tau} &\;=\; -\,\frac{\p H}{\p g}\,,\qquad\quad  
\frac{{\rm d}g}{{\rm d}\tau} \;=\; \frac{\p H}{\p L}\,;\qquad\quad
\frac{{\rm d}L_z}{{\rm d}\tau} \;=\; -\,\frac{\p H}{\p h}\,,\qquad\quad  
\frac{{\rm d}h}{{\rm d}\tau} \;=\; \frac{\p H}{\p L_z}\,.
\label{eom-ring}
\end{align}

\item {$F(\scrr, \tau)$ obeys the secular CBE:}  
\beq
\frac{{\rm d}F}{{\rm d}\tau} \;\equiv\;
\frac{\p F}{\p \tau} \;+\; \left[\,F\,,\,H\,\right] \;=\; 0\,,
\label{cbe-ring}
\eeq
{where $[\;,\;]$ is the 4--dim Poisson Bracket,} 
\beq
\left[\,\chi_1\,,\,\chi_2\,\right] \;\stackrel{{\rm def}}{=}\; 
\left(\frac{\p \chi_1}{\p g}\frac{\p \chi_2}{\p L} -
\frac{\p \chi_1}{\p L}\frac{\p \chi_2}{\p g}\right) \,+\, 
\left(\frac{\p \chi_1}{\p h}\frac{\p \chi_2}{\p L_z} -
\frac{\p \chi_1}{\p L_z}\frac{\p \chi_2}{\p h}\right)\,.
\label{pbdel4}
\eeq

\item {Secular collisionless equilibria $F= F_0(\scrr)$ are stationary solutions of the secular CBE eqn.(\ref{cbe-ring}) and satisfy, 
$\left[\,F_0(\scrr)\,,\,H_0(\scrr)\,\right] = 0\,$. These can be constructed by a secular Jeans' theorem which states that $F_0$ is a function of 
$\scrr$ only through the time-independent integrals of motion of $H_0(\scrr)$, and any (positive and normalized) function of the time-independent integrals of $H_0(\scrr)$ is a stationary solution of equation~(\ref{cbe-ring}).}
\end{itemize}

{\subsection{The cusp-disc system}}

{We are now in a position to formulate our problem in terms of the above description of secular collisionless dynamics.}

\medskip
\noindent
{\bf The unperturbed cusp:}
The secular DF for the spherical unperturbed cusp is
\beq
F_0(I,L) \;=\; 2\pi \, f_{\rm c}(E_{\rm k}, L) \;=\; \frac{A \,(G\mdot)^{2\, m} \, L^n}{2^m \, I^{2 \, m}}\,, 
\label{cusp-df-I}
\eeq
where we have used equation~(\ref{cusp-df}). {The corresponding (scaled) orbit-averaged potential, $\Phi_{\rm c}(I, L)$, 
is related to $F_0$ through equation~(\ref{phislow-r}), but we do not need 
to use this; it is easier to orbit-average equation~(\ref{cusp-pot}).} 
Then we get $\,\Phi_{\rm c}(I, L) = (\mdot/M_{\rm c})\oint \varphi_{\rm c}(r)\,{\rmd w}/{2\pi}\,$, is proportional to a hypergeometric function, but the following approximate expression will suffice for our purposes:\footnote{Both the exact expression and the approximation are given in equations~(4.81) and (4.82) of \citet{mer13}.}  
\beq
\Phi_{\rm c}(I,L) \;=\; \frac{G\mdot}{(2-\gamma) \, \rc} \left( \frac{a}{\rc} \right)^{2-\gamma} (1+\alpha_{\gamma} \, e^2)\,,\qquad\mbox{where}\quad \alpha_{\gamma} \;=\; \frac{ 2^{3-\gamma} \, \Gamma{(\frac{7}{2}-\gamma)}}{ \sqrt{\pi} \, \Gamma{(4-\gamma)} } \,-\, 1\,. 
\label{phic-avg}
\eeq
This formula is exact for $\gamma = 1$, and a good approximation 
for our fiducial value, $\gamma = 5/4$. $\,\Phi_{\rm c}(I,L)$ acts as 
the Hamiltonian for secular dynamics so the apse precession frequency, 
$\rmd g/\rmd\tau = \Omega_{\rm c}(I,L)$, is:
\beq
\Omega_{\rm c}(I,L) \;=\; \frac{\partial \Phi_{\rm c}}{\partial L} \;=\; -\,\frac{2 \, \alpha_{\gamma}}{2-\gamma}\, \Omega_{\rm kep}(r_{\rm c})\, \frac{I^{3-2\, \gamma} }{ (G\mdot\,r_{\rm c})^{\frac{3}{2}-\gamma}}\,\frac{L}{I}\,,
\label{omega_c}
\eeq
where $\Omega_{\rm kep}(r_{\rm c}) = (G\mdot/r_c^3)^{1/2}\,$ is the Kepler frequency for an orbit of semi--major axis $\rc$. Since $\Omega_{\rm c} \propto -a^{(3/2 - \gamma)}\,\sqrt{1-e^2\,}$, the (retrograde) apse precession is fastest for near-circular orbits and and slowest for highly eccentric orbits. Moreover for $\gamma < 3/2$, which is of interest to us, orbits of smaller $a$ precess slower.  

\medskip
\noindent
{\bf Orbit-averaged disc perturbation}: $\,\Phi_{\rm d}(I, L, L_z, g, \tau) = (\mdot/M_{\rm c})\oint \varphi_{\rm d}(r, \theta, t)\,{\rmd w}/{2\pi}\,$ 
can be written in terms of Elliptic integrals for the potential of
equation~(\ref{disc-pot}), as given in Appendix~A. The following 
approximation, which is convenient for calculations, has a maximum fractional error $\lesssim 2 \%\,$:
\begin{align}
\Phi_{\rm d} \;=\; 
\frac{16\,G\mdot}{11\pi\,r_{\rm c}}\mu(\tau)\sqrt{\frac{r_{\rm d}}{a}} 
\Bigg[&-\frac{9}{100}\sqrt{1+e}\;{\cal E}\!(k)\left(33 + \frac{\sin^2{i}}{2}\right)  + \frac{\sin{i}}{2}\left(1 + a_0e^2 + b_0e^4 + c_0e^6\right) 
\Bigg.\nonumber\\[1ex]
\Bigg.&-\left(\frac{\lambda}{2}\sin{i} - \frac{9}{100}\sin^2{i}\right)\left(a_te^2 + b_te^4 + c_te^6\right)\cos{2g}\,\Bigg]\,,
\label{phid-avg}
\end{align}
where $k = \sqrt{2e/(1+e)}$, $\;{\cal E}\!(k)$ is the complete elliptic integral of second kind defined in equation~(\ref{ell-int-2k}), and $a_0 = - 0.0742572,  \,b_0=0.0417887, \,c_0 =- 0.0672152, \,\lambda = 0.848835, \,a_t = 0.495367, \,b_t=-0.492259, \,c_t =0.703998$. Here $\mu(\tau) = \left[M_{\rm d}(\tau) \,\rc/M_{\rm c}\,r_{\rm d}\right]$ is a time-dependent small parameter characterising the strength of the disc perturbation relative to the cusp: $\mu(\tau) \to 0$ as $\tau\to -\infty$ and $\mu$ takes its largest value of $0.1$ when $M_{\rm d} = 10^5~\msun$.

\medskip
\noindent
{\bf Secular evolution of the cusp DF}: The spherical cusp DF of equation~(\ref{cusp-df-I}) responds to the time-dependent, axisymmetric disc potential of equation~(\ref{phid-avg}). The DF of the axisymmetrically deforming cusp must be independent of the nodal longitude $h$, and takes the general form, $F(I, L, L_z, g,\tau)$. Let $\Phi(I, L, L_z, g,\tau)$ be the (scaled) self-gravitational potential, which is related to $F$ through equation~(\ref{phislow-r}). The secular Hamiltonian is, 
\beq
H(I, L, L_z, g,\tau) \;=\; \Phi(I, L, L_z, g,\tau) \;+\; 
\Phi_{\rm d}(I, L, L_z, g, \tau)\,.
\label{ham}
\eeq
Since both $F$ and $H$ are independent of $h$, the CBE of equation~(\ref{cbe-ring}) simplifies to,
\beq
\frac{\partial F}{\partial \tau} \;+\; \frac{\partial H}{\partial L}\frac{\partial F}{\partial g}  \;-\; \frac{\partial H}{\partial g}
\frac{\partial F}{\partial L}  \;=\; 0\,.
\label{cbe}
\eeq
Both $I= \sqrt{G\mdot a\,}$ and $L_z = I\sqrt{1-e^2\,}\,\cos{i}$ are secular integrals of motion, even though $H$ is time-dependent. If $H$ were time-independent, it is itself a third integral of motion; in contrast to un-averaged stellar dynamics, all time-independent, axisymmetric secular dynamics is integrable \citep{st99}. Then the secular Jeans theorem \citep{st16} implies that a steady state $F$ must be function of $(I, L_z, H)$. We need to solve the problem for an adiabatically varying $H$.

\section{Adiabatic response of the stellar cusp}

The time-dependence of $H$ is driven by disc growth over times, $T_{\rm grow} \gtrsim 10^7~\mbox{yr}$, that are much longer than $\ts \lesssim 2\times 10^5~\mbox{yr}$. In this case $H$ is not conserved, but the principle of 
adiabatic invariance can be used to calculate a new action, $J = \oint L(H, I, L_z, g, \tau)\,{\rmd g}/2\pi\,$, that is conserved for orbits that are
far from a separatrix, and undergoes a probabilistic change which
can be calculated for orbits encountering a separatrix \citep{gp66,hen82}; the corresponding evolution of the DF was worked out in \citet{st96} --- 
see \S~4.2 for a more detailed discussion of these points. The non-linear, axisymmetric, adiabatic response is an integrable and solvable problem.
 We derive an explicit formula for the linear response of the DF, due to the growing disc potential while neglecting the change in the cusp potential, as discussed below. This is used in the next section to calculate density deformation. Then we study orbital structure: this provides a physical interpretation of the linear deformation, clarifies the limits of linear theory and sets the stage for the non-linear theory of adiabatic deformation.

\subsection{Linear adiabatic response}

The unperturbed cusp has DF $F_0(I, L)$ and Hamiltonian 
$H_0 = \Phi_{\rm c}(I, L)$. As the disc grows the cusp DF is $F = F_0(I, L) + F_1(I, L, L_z, g,\tau)$, with the corresponding new Hamiltonian $H = H_0 + H_1$ where $H_1 = \Phi_{\rm d}(I, L, L_z, g,\tau) + 
\Phi_1(I, L, L_z, g,\tau)$. Here $\Phi_1$ is the (scaled) self-gravitational potential due to $F_1$, and related to it through the Poisson integral of equation~(\ref{phislow-r}): 
\beq
\Phi_1(I, L, L_z, g,\tau) \;=\; \int F_1(I, L, L_z, g,\tau)\,\Psi(\scrr, \scrr')\,{\rm d}\scrr'\,.
\label{poisson}
\eeq
From the discussion of time scales in \S~2.3, we expect that disc perturbation is small for $a \gtrsim 0.2~\mbox{pc}$. Substituting for $F$ and $H$ in the CBE of equation~(\ref{cbe}), and keeping only terms linear in the small quantities, $\left\{F_1, \Phi_{\rm d}, \Phi_1\right\}$, we obtain the linearised collisionless Boltzmann equation (LCBE) governing the evolution of $F_1\,$:
\beq
\frac{\partial F_1}{\partial \tau} \;+\; \Omega_{\rm c}(I, L)\, \frac{\partial F_1}{\partial g}  \;=\; \frac{\partial F_0}{\partial L}  \frac{\partial}{\partial g}\left\{\Phi_{\rm d} + \Phi_1\right\}.
\label{lcbe}
\eeq
The price to be paid for linearization is that we will not be able to describe capture into resonance (which is discussed later in \S~4.2). 

Since $\Phi_1$ is given as an integral over $F_1$, the LCBE
is a linear integro-differential equation for the unknown $F_1$. Calculating even this linear response requires substantial numerical computations. For a 
first cut at the problem we proceed by dropping $\Phi_1$ (the likely 
effect of this would be to underestimate the response of the cusp). Then the right side of equation~(\ref{lcbe}), $(\partial \Phi_{\rm d}/\partial g)$, represents only the known driving due to the disc, and the LCBE reduces to a linear partial differential equation. Further simplification occurs because
of the adiabaticity of the problem, which was established in \S~2.3: the first term on the left side, $(\partial F_1/\partial \tau)$, is smaller than the second term, $\Omega_{\rm c}(\partial F_1/\partial g)$, by a factor $(T_{\rm pr}^{\rm c}/T_{\rm grow}) \sim 2\times 10^{-2}$. Hence, dropping $\partial F_1/\partial\tau$, we can integrate over $g$ to find $F_1$. \footnote{Since $\left|\Omega_{\rm c}\right| \propto a^{(3/2 - \gamma)}\,\sqrt{1-e^2\,}$ decreases as $a$ decreases (for $\gamma < 3/2$), and $e$ increases, this assumption is not valid for small and/or highly eccentric orbits. But we need to account for non-linear effects long before we face this limitation of the adiabatic approximation in the linear theory itself. This is discussed later in this section.} The physical solution cannot have a $g$-independent part because such a deformation is not allowed through collisionless, secular Hamiltonian deformations in phase space. Therefore
\beq
F_1(I, L, L_z, g,\tau) \;=\; \frac{1}{\Omega_{\rm c}(I, L)}\frac{\partial F_0}{\partial L}\left[\,\Phi_{\rm d} \;-\; \left<\Phi_{\rm d}\right>_g\,
\right]\,,
\label{f1-phys}
\eeq
where $\left<\Phi_{\rm d}\right>_g = \oint\Phi_{\rm d}\,\rmd g/2\pi\,$.
Using the purely $g$-dependent part on the right side of equation~(\ref{phid-avg}), together with equations~(\ref{omega_c}) and (\ref{cusp-df-I}),   
we obtain the following explicit expression:
\begin{align}
F_1 &\;=\;
\frac{D(\tau)}{(G\mdot\rc)^{3/2}}\,\frac{\rc}{a}\,(1-e^2)^{\left(\frac{n}{2} -1\right)}\left(a_te^2 + b_te^4 +c_te^6 \right)\left(\frac{\lambda}{2} \sin{i} - \frac{9}{100} \sin^2{i}\right)\cos{2g}\,,\nonumber\\[1em] 
&\mbox{where}\qquad D(\tau) \;=\; \frac{4n\,(2-\gamma)(3-\gamma)}{11\pi^2\, \alpha_{\gamma}\,2^{(\gamma +n)}\,B_{\left(\frac{n}{2} + 1,\frac{1}{2}\right)}\,B_{\left(m+1,\frac{n+3}{2}\right)}}\,\sqrt{\frac{r_{\rm d}}{\rc}} 
\;\mu(\tau)\,.
\label{f1-orb}
\end{align}
The secular linear deformation has been written in terms of physical variables, instead of Delaunay variables, so we can read-off its general properties:
\begin{itemize}
\item[{\bf 1.}] $\,F_1 \propto a^{-1}$ is independent of the cusp power-law index because $\gamma$ cancels out in the ratio, $\Omega_{\rm c}^{-1}\,(\partial F_0/\partial L)$, in equation~(\ref{f1-phys}).  The magnitude of $F_1$ increases with decreasing $a$ because the perturbing gas density rises steeply at small radii. Linear theory requires that $\vert F_1\vert \ll \vert F_0\vert \propto a^{3/2 - \gamma}$, so applies at small $a$ only when $\gamma > 5/2$. For the shallow cusp we consider, $\gamma \approx 5/4$, equation~(\ref{f1-orb}) would not correctly represent the perturbation at small $a$.

\item[{\bf 2.}] The magnitude of $F_1$ is an increasing function of the inclination, $i\,$, because $F_1$ is proportional to the $g$-dependent
part of the disc potential, whose effect increases with inclination. 

\item[{\bf 3.}] For $n \leq 2$, the magnitude of $F_1$ is an increasing function of the eccentricity, $e\,$. For $n > 2$ orbits with intermediate values of $e$ contribute the most, because the unperturbed cusp has very tangentially biased velocity dispersions.

\item[{\bf 4.}] Since $F_1 \propto \cos{2g}\,$ it is positive/negative for orbits whose angles between their lines of apses and nodes is lesser/greater than $45^{\circ}$. $F_1$ is positive and maximum for $g = \left(0^{\circ}, 180^{\circ}\right)$, and negative and minimum for $g = \left(90^{\circ}, 270^{\circ}\right)$. 
\end{itemize}
Of the four properties the first three pertain to the magnitude of $F_1$.
The fourth item alone determines the sign of $F_1$, and hence the flattening of the cusp. In order to understand this physically it is necessary to 
work out the broad characteristics of the individual orbits making up the stellar system. This also enables an appreciation of what is involved
in calculating non-linear, adiabatic response.

\subsection{Orbital structure and non-linear theory}

The Hamiltonian governing orbital structure is $H(I, L, L_z, g,\tau) = \Phi_{\rm c} + \Phi_{\rm d}$. Using equations~(\ref{phic-avg}) and (\ref{phid-avg}) we have:
\begin{align}
H &\;=\; \frac{G \, \mdot}{r_{\rm c}} \Bigg[\frac{1}{(2-\gamma) } \left( \frac{a}{\rc} \right)^{2-\gamma} (1+\alpha_{\gamma} \, e^2) + \frac{16\,\mu(\tau)}{11\,\pi}  \sqrt{\frac{r_{\rm d}}{a}} \left\{ -\frac{9}{100} \sqrt{1+e} \; {\cal E}\!\left( k \right) \, \left( 33 + \frac{\sin^2{i}}{2} \right)\right.\Bigg.\nonumber\\[1ex] 
&\quad\Bigg.\left.+\; \frac{\sin{i}}{2} \, \left(1 + a_0e^2 + b_0e^4 + c_0e^6\right) - \left( \frac{\lambda}{2} \sin{i} - \frac{9}{100} \sin^2{i} \right) \, (a_te^2 + b_te^4 + c_te^6) \, \cos{2g} \right\} \Bigg]\,.
\end{align}
As we discussed at the end of \S~3, this time-dependent Hamiltonian always
has two integrals of motion, $I= \sqrt{G\mdot a\,}$ and $L_z = I\sqrt{1-e^2\,}\,\cos{i}$. Therefore the eccentricity and inclination execute coupled oscillations: when $e$ increases $i$ decreases, while $a = \mbox{constant}$.
In order to say more about orbits we need some information on the time-dependence of $H$, which arises through the parameter $\mu(\tau)$.

\medskip
\noindent
{\bf `Time-frozen' Hamiltonian:}
Were $\mu(\tau) = \mbox{constant}$, then $H$ would be time-independent, and is itself the third integral of motion. Orbital dynamics can be visualised by first fixing some values of $(I, L_z)$, and drawing isocontours of $H$ in the $(L, g)$ phase plane, for $L \geq \left|L_z\right|$. For $\mu = 0$ we have $H = \Phi_{\rm c}(I, L)$, so the isocontours are just $L= \mbox{constant}$ horizontal lines. For $\mu \neq 0$ the isocontours have a more complicated topology: these are displayed in Figure~2 for $\mu = 0.1$ (its maximal value), $\,a = 0.5~\mbox{pc}$ and two different values of $L_z$. The orbital structure shares the following generic features of secular dynamics in time-independent, axisymmetric potentials around a MBH \citep{ss00, mer13}:
\begin{itemize}

\item[] \emph{Circulating orbits}, for which $g$ advances by $2\pi$ over 
one period. These can be thought of as perturbations of the $L = \mbox{constant}$ orbits of the $\mu = 0$ case, exhibiting periodic 
oscillations of both $L$ and $g$. The perturbations need not necessarily be small, but they are small enough so that the basic topology of the orbit
remains unchanged.  

\item[] \emph{Librating orbits}, for which $g$ librates periodically about 
$g = (\pi/2, 3\pi/2)$. These populate two `islands' parented by two
elliptic fixed point orbits (marked by the dots), which correspond to Kepler ellipses of fixed $(a, e, i, g)$ whose nodes precess at a steady rate.

\item[] \emph{Two Separatrix} orbits (dashed lines) that meet at the hyperbolic fixed points at $g = (0, \pi)$. These partition the phase plane into circulating and librating orbits. The period of a separatrix orbit is infinite, as apse precession slows down terminally near the fixed points.
\end{itemize}

\begin{figure}  
\begin{subfigure}{0.5\textwidth}
\centering
\includegraphics[width=0.9 \textwidth,trim={3cm 0cm 2cm 0.5cm}]{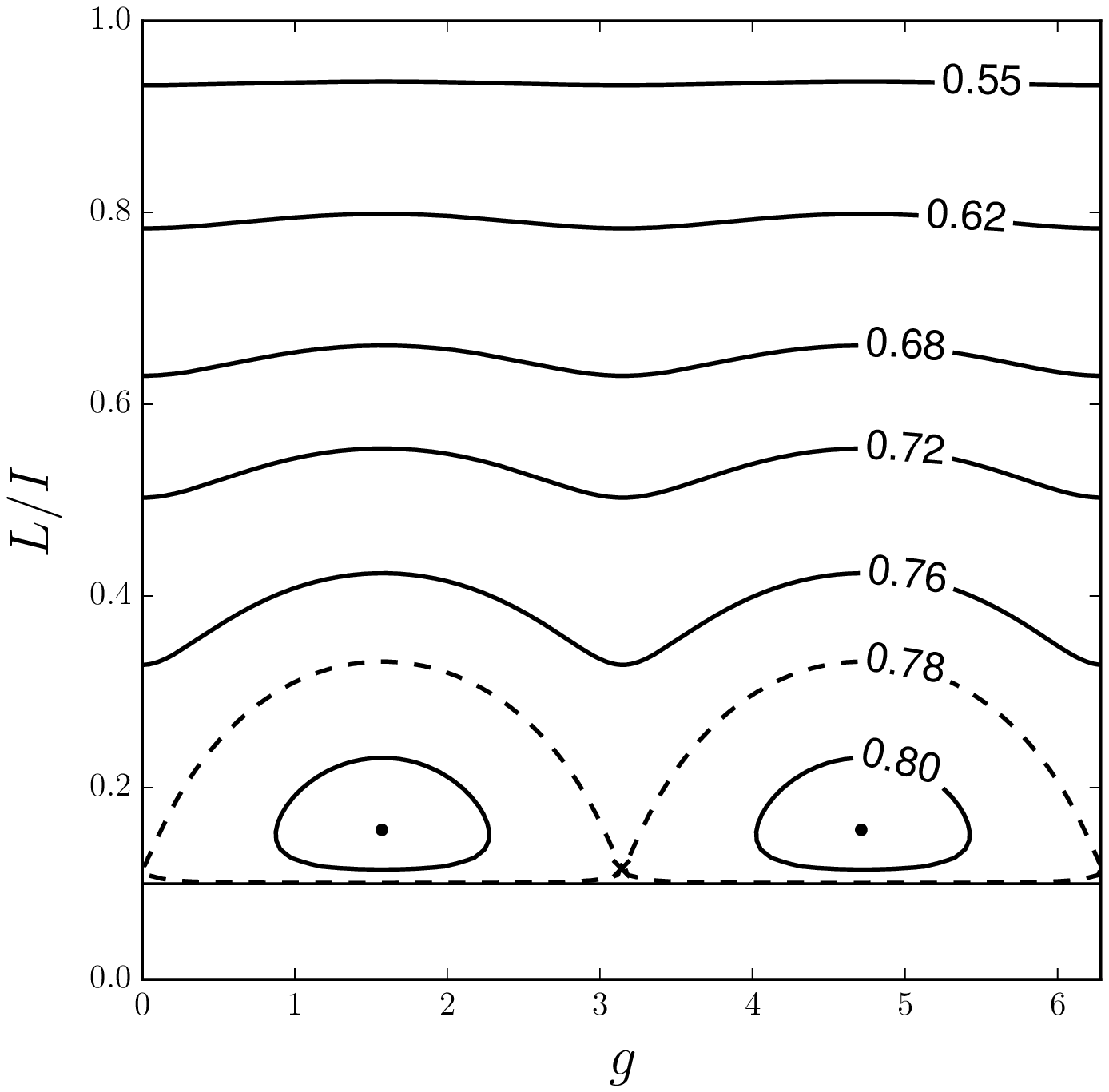}
\subcaption{$L_z/I = 0.1$}
\end{subfigure}
\hfill
\begin{subfigure}{0.5\textwidth}
\centering
\includegraphics[width=0.9 \textwidth,trim={3cm 0cm 2cm 0.5cm}]{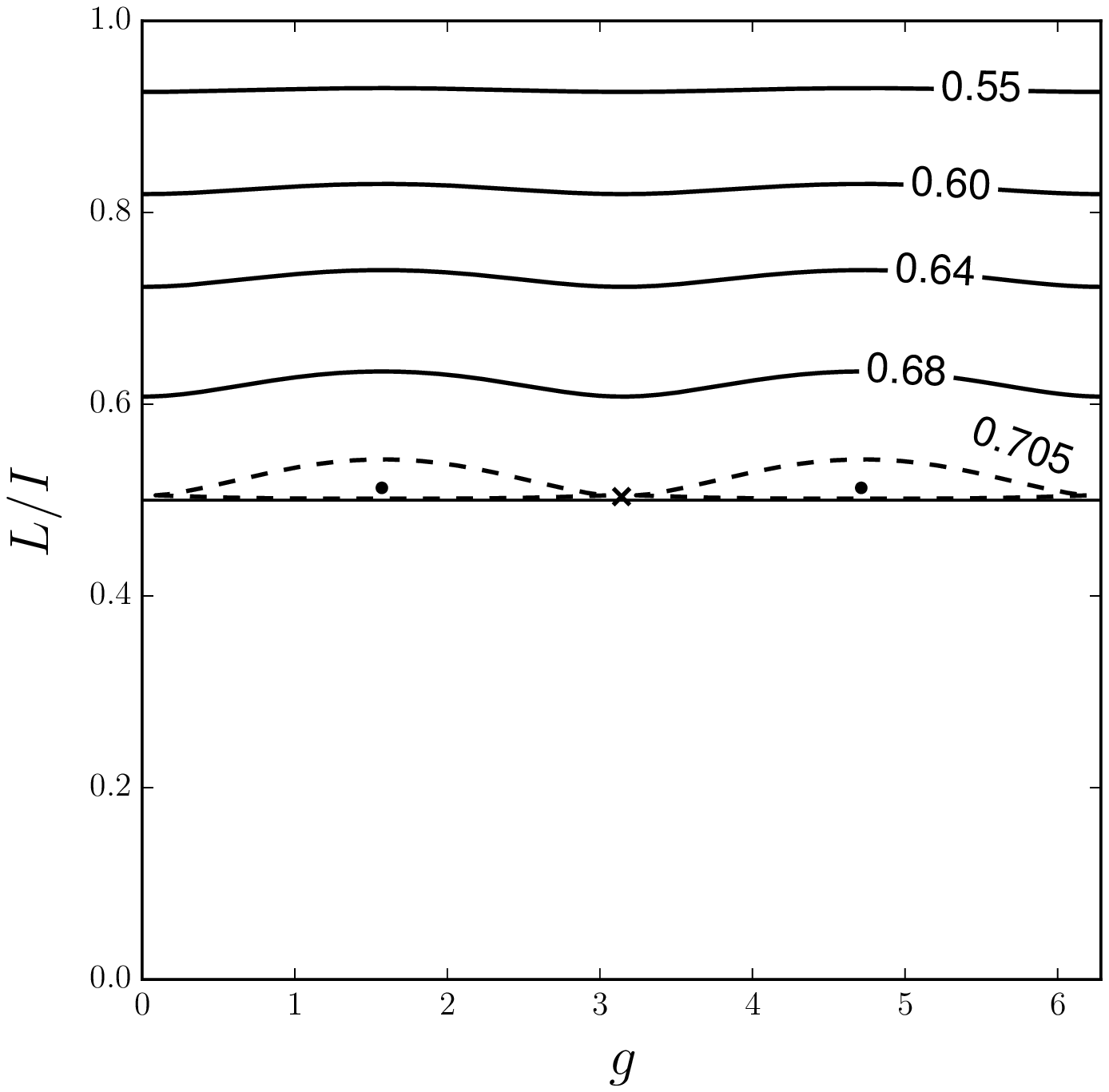}
\subcaption{$L_z/I = 0.5$}
\end{subfigure}
\caption{\emph{Isocontours of $H(I,L,L_z,g)$ in the $(L, g)$ phase plane}, in units of $G \mdot/\rc$, for $\mu = 0.1$ and a = $0.5$ pc. The exact expressions for $\Phi_{\rm c}$, given in equation~(4.81) of \citet{mer13}, and $\Phi_{\rm d}$, given in equation~(\ref{avg-phi-d-exact}), have been used.}
\label{H-contours}
\end{figure}

\medskip
\noindent
{\bf Adiabatically varying Hamiltonian:}
When $\mu(\tau)$ varies slowly with time, $H$ is no longer an integral of motion. At early times $\mu \to 0\,$ so $\,H \to \Phi_{\rm c}(I,L)$, which
is just the unperturbed cusp. All orbits circulate at constant $L$, corresponding to retrograde apse precession at the constant rate $\Omega_{\rm c}$. As $\mu(\tau)$ increases two islands appear around the elliptic fixed points, together with their separatrices. As $\mu(\tau)$ increases the separatrices expand and the islands grow until their areas attain a maximum when $\mu = 0.1\,$. There are two cases to consider:

{\bf (1) \emph{Adiabatic invariance and linear theory}:}
For circulating orbits that do not ever encounter the growing separatrices, 
$\mu(\tau)$ may be considered to be slowly varying. Then $J = \oint L(H, I, L_z, g, \tau)\,{\rmd g}/2\pi\,$ is an adiabatic invariant, so we have three secular integrals of motion, $(I, L_z, J)$. The secular Jeans theorem
implies that the full, non-linear DF is of the form $F(I, L_z, J)$. 
The linear response calculation of \S~4.1 is a particular case, valid for those circulating orbits that remain close to an unperturbed $L = \mbox{constant}$ orbit. In this case $F = F_0(I, L) + F_1(I, L, L_z, g, \tau)$, where $F_0$ and $F_1$ are given in equations~(\ref{cusp-df}) and (\ref{f1-orb}). We can now understand the general form of $F_1$, by following 
individual circulating orbits. 

\begin{figure}
\centering
\includegraphics[width=0.7\textwidth]{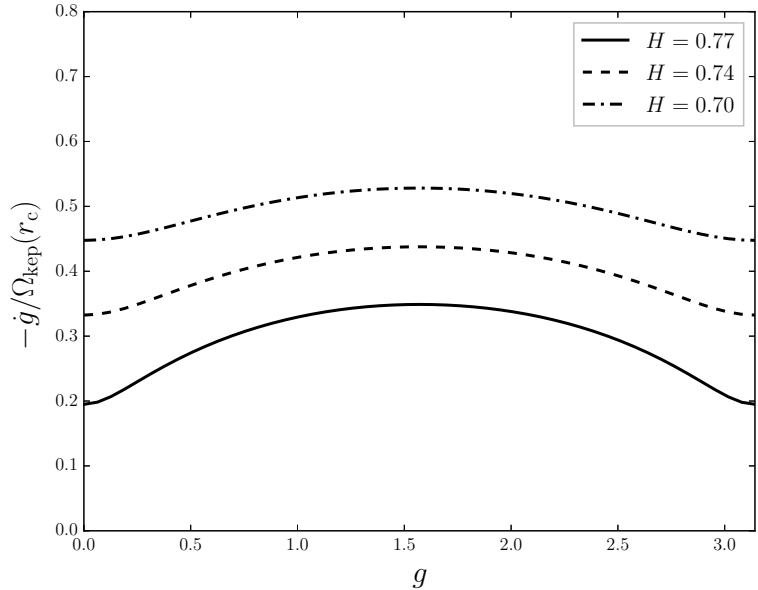}
\caption{\emph{Apse precession rates for three circulating orbits} in the 
phase plane of Figure~\ref{H-contours}a, for $H \,=\, 0.70,\;0.74,\;0.77$.} \label{orbits}
\end{figure}

From Figure~\ref{H-contours} and the conservation of $L_z = L\cos{i}$, we 
see that both $L$ and $i$ take their smallest value at $g = \left(0^{\circ}, 180^{\circ}\right)$, and largest value at $g = \left(90^{\circ}, 270^{\circ}\right)$. Figure~\ref{orbits} shows the (retrograde) apse precession rate, $\dot{g} = \partial H/\partial L$, as a function of $g$, for three  circulating orbits taken from the left panel of Figure~\ref{H-contours}. 
Apse precession is slowest at $g = \left(0^{\circ}, 180^{\circ}\right)$, and fastest at $g = \left(90^{\circ}, 270^{\circ}\right)$. Since the orbit 
spends the most time where it precesses slowest, we expect a positive
perturbation to the DF near $g = \left(0^{\circ}, 180^{\circ}\right)$,
when the orbit also attains its maximum eccentricity and minimum inclination. Precisely the opposite behaviour obtains near $g = \left(90^{\circ}, 270^{\circ}\right)$. All of these contribute to an over-density in the perturbation close to the disc plane, and an under-density away from the disc plane, thereby flattening the cusp. Indeed the density deformation $\rho_1$, shown in Figure~\ref{rho1}, has this expected form. 

{\bf (2) \emph{Adiabatic capture and non-linear theory}:} When a circulating orbit encounters one of the growing separatrices, it will be captured into the respective island and become a librating orbit. We now discuss the generic situation, which includes cases when one or both separatrices shrink.

Adiabatic invariance is broken in the vicinity of a time-dependent separatrix, both on the librating and circulating sides. This is because the orbital periods are formally infinite on the separatrices, and there is a band of actions around the separatrices for which the orbital periods are longer than the time of variation of the self-consistent Hamiltonian. This band, which includes the unstable fixed points, is very narrow in the adiabatic limit. But for orbits within it, the movement of the separatrices is not slow, and the dynamics within the band is chaotic because the orbit--separatrix encounter is very sensitive to the phase of the encounter. The behaviour of the orbit has been described in probabilistic terms in the planetary dynamics literature \citep{gp66,hen82}; i.e. in terms of the probabilities of capture into, or escape from the islands of libration. \citet{st96} reconsidered this general problem in terms of the collisionless behaviour of a distribution of particles, and showed that the capture/escape probabilities can be calculated, without doing the detailed non-linear dynamics of the encounter of an orbit with a separatrix. We note their main results, and discuss it in the context of our problem:
  
\begin{itemize}
\item Let $f$ be the fine-grained DF of the particles that obeys the CBE, whose Hamiltonian (which could be self-consistent or not) allows for a resonant island bounded by separatrices, which distort over time scales much larger than generic orbital periods (by generic we mean 
orbits that do not lie in the narrow band discussed above). Even if $f$ was a smooth function to begin with, the chaotic orbit-separatrix encounter discussed above results in the post-encounter DF acquiring extremely fine-grained structure within the narrow band around the separatrix. 

\item We begin by noting that, at any given time, the
band around the separatrices is very narrow. Then the fine-grained structure is essentially reflected in a rapid dependence of $f$ as a function of 
the instantaneous angle variable. Hence it seems natural to introduce a coarse-grained DF, $\bar{f}$, which equals $f$ averaged over the instantaneous angle variable.

\item From the single principle of conservation of the total mass in the coarse-grained DF, $\bar{f}$, \citet{st96} derived the 
evolution of $\bar{f}$ in phase space at any given time: (i) Away from the separatrices $\bar{f}$ retains its adiabatic invariant form, for both circulating and librating orbits; (ii) In the immediate vicinity of the separatrices, $\bar{f}$ undergoes changes, as listed in Table~1 of their paper. These rules automatically provide the classical expressions for capture probabilities, derived in planetary dynamics, so the coarse-grained description indeed gives correct results. 

\item The rules for $\bar{f}$ around the separatrices are such that all entropy (or ${\cal H}$) functions associated with it grow in time (in contrast all entropy functions computed with respect to the fine-grained DF, $f$, are conserved during collisionless evolution). Hence the coarse-grained evolution is both mixing and irreversible, which should not be surprising because the nonlinear dynamics within the band around the separatrices is chaotic.
\end{itemize}

In the context of the cusp-disc problem studied in this paper, the islands grow monotonically from vanishingly small sizes in the distant past. Hence every librating orbit was once a circulating orbit that was captured by the growing separatrices. Since the DF inside the islands is built up over time by capturing circulating orbits, the DF for the librating orbits depends on the entire time evolution of the system, in contrast to the case discussed above when $J$ was conserved. The secular adiabatic evolution of an axisymmetric system --- even when the self-gravity of the perturbation is included --- is an integrable problem. So the full non-linear problem, with application of the rules from \citet{st96}, can be 
computed in a definite manner, but this is beyond the scope of this paper.

\FloatBarrier
\section{Spheroidal flattening of the cusp}

Here we compute the deformation of the three dimensional density and 
the surface density, as seen from different viewing angles. The density perturbation can be calculated by integrating  $F_1$ of equation~(\ref{f1-orb}) over velocity space. This can be carried through analytically
(see Appendix~B), and the result is this simple formula:  
\begin{align}
\rho_1(r,\theta,\tau) &\;=\; \frac{\Mc}{2\pi}\int F_1(I, L, L_z, g, \tau)\, 
\rmd\bfu \;=\; \frac{3-\gamma}{4\pi}\,C_{n,\gamma}(\tau)\,\frac{M_{\rm c}}{\rc^3} \left(\frac{\rc}{r_{}}\right)^{\!\frac{5}{2}}\,\Theta(\theta)\,,  
\nonumber\\[1em]
\mbox{where}\qquad &\Theta(\theta) \;=\; \frac{\lambda}{2\pi}\left[\,{\cal E}(\sin{\theta}) \,-\, 2 \cos^2\theta \,{\cal K}(\sin{\theta})
\right] \;-\; \frac{9}{400}(1 \,-\,3\cos^2{\theta})\,;\nonumber\\[1em] 
&C_{n,\gamma}(\tau) \;=\; \frac{16n\,(2-\gamma)\,{\cal B}(n,\gamma)}{11\pi \, 2^{\left(\gamma-\frac{1}{2}\right)} \, \alpha_{\gamma}}\, \sqrt{\frac{r_{\rm d}}{\rc}}\,\mu(\tau)\,. 
\label{rho1-final}
\end{align}
Here ${\cal B}(n,\gamma)$ is a function of the indices, $(n, \gamma)$, 
of the unperturbed spherical cusp, as given in equation~(\ref{rho1-fin-app}).
It should be noted that the dependence of $\rho_1$ on $r$ and $\theta$ is independent of $(n, \gamma)$.
 
This expression for $\rho_1$ is valid only when the $F_1$ of equation~(\ref{f1-orb}) is a reasonable approximation. This would be true 
for many of the circulating orbits of Figure~\ref{H-contours} but not
for the librating orbits that are trapped in the islands, as discussed
in the previous section. For any $(I, L_z)$ the librating orbits occur for the lowest values of $L$, so linear theory cannot be expected to work well
when the unperturbed cusp has radially anisotropic velocity dispersions. 
But the GC cusp is probably tangentially anisotropic, with $\beta \approx -1/4$ for  $r < 2~\mbox{pc}$ \citep{feld17}, so we can expect the linear theory result of equation~(\ref{rho1-final}) to be a useful first approximation. 

\begin{figure}  
\begin{subfigure}{0.5\textwidth}
\centering
\includegraphics[width=0.9 \textwidth,trim={3cm 0cm 2cm 0cm}]{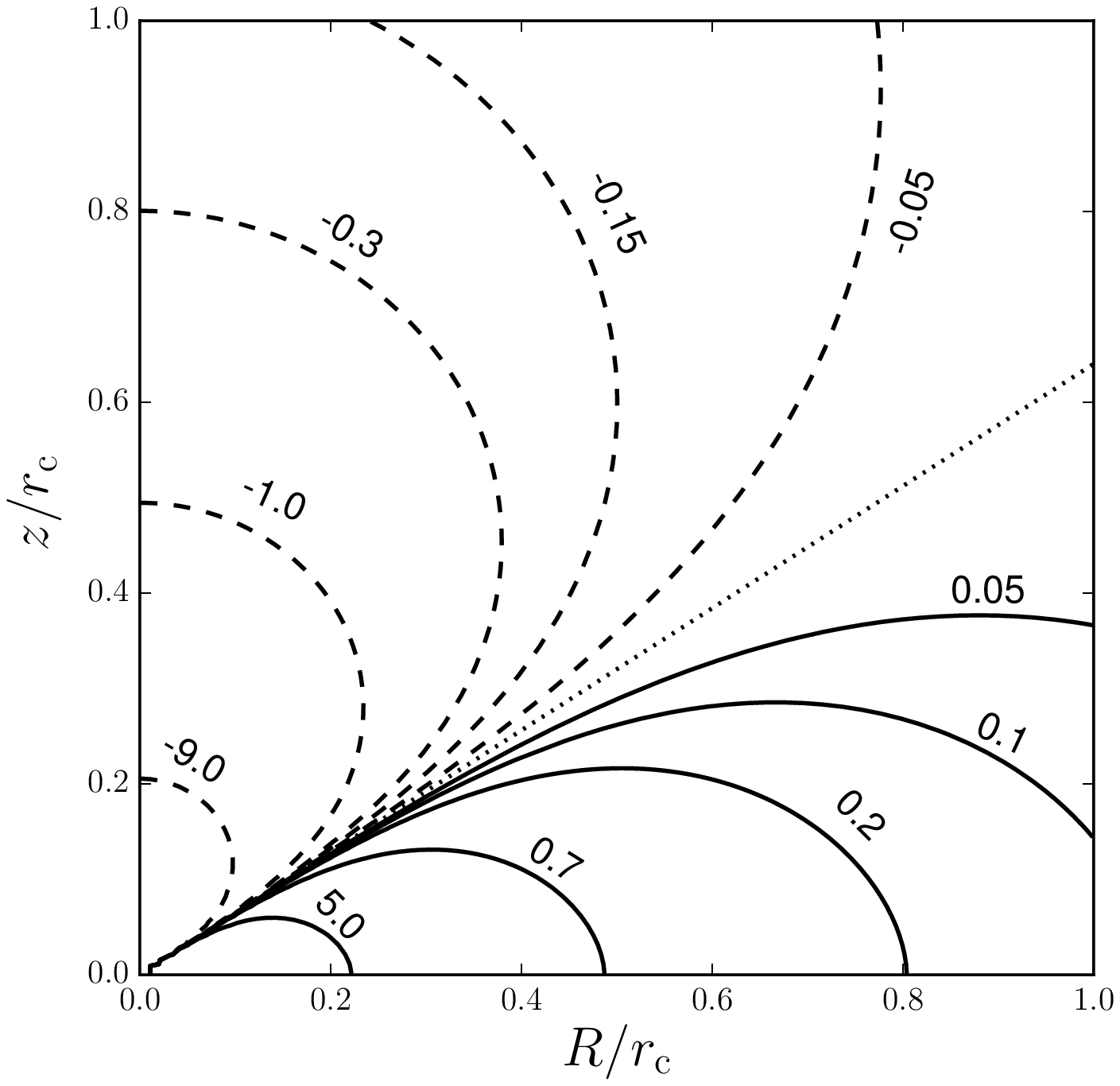}
\subcaption{Density perturbation, $\rho_1$, in units of $ 10^{-2} \,M_{\rm c}/\rc^3$.}
\label{rho1}
\end{subfigure}
\hfill
\begin{subfigure}{0.5\textwidth}
\centering
\includegraphics[width=0.9 \textwidth,trim={3cm 0cm 2cm 0cm}]{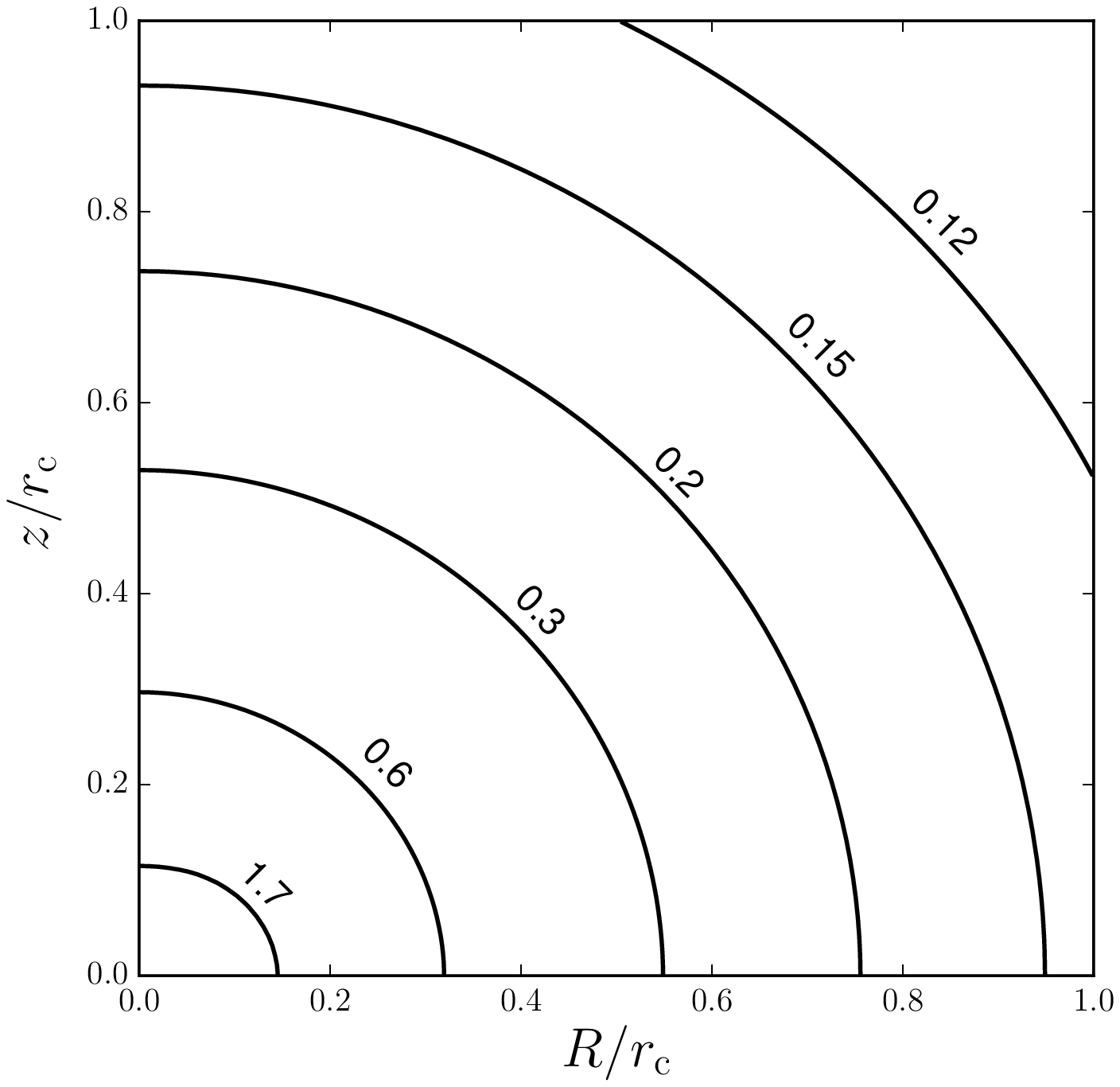}
\subcaption{Total density, $\rho$, in units of in $M_{\rm c}/\rc^3$.}
\label{rho}
\end{subfigure}
\caption{\emph{Cusp deformation}: Isocontours of three dimensional densities, for $\gamma = 5/4$ and $n=1/2$. [Left Panel] Solid curves are for $\rho_1 > 0$, and dashed curves are for $\rho_1 < 0$; the dotted straight line at
$\theta = 57.37^{\circ}$ is for $\rho_1 = 0$. [Right Panel] Isocontours of 
the total density, $\rho$, showing an oblate spheroidal deformation.}
\end{figure}

\begin{figure}  
\begin{subfigure}{0.5\textwidth}
\centering
\includegraphics[width=0.9\textwidth,trim={3cm 0cm 2cm 0cm}]{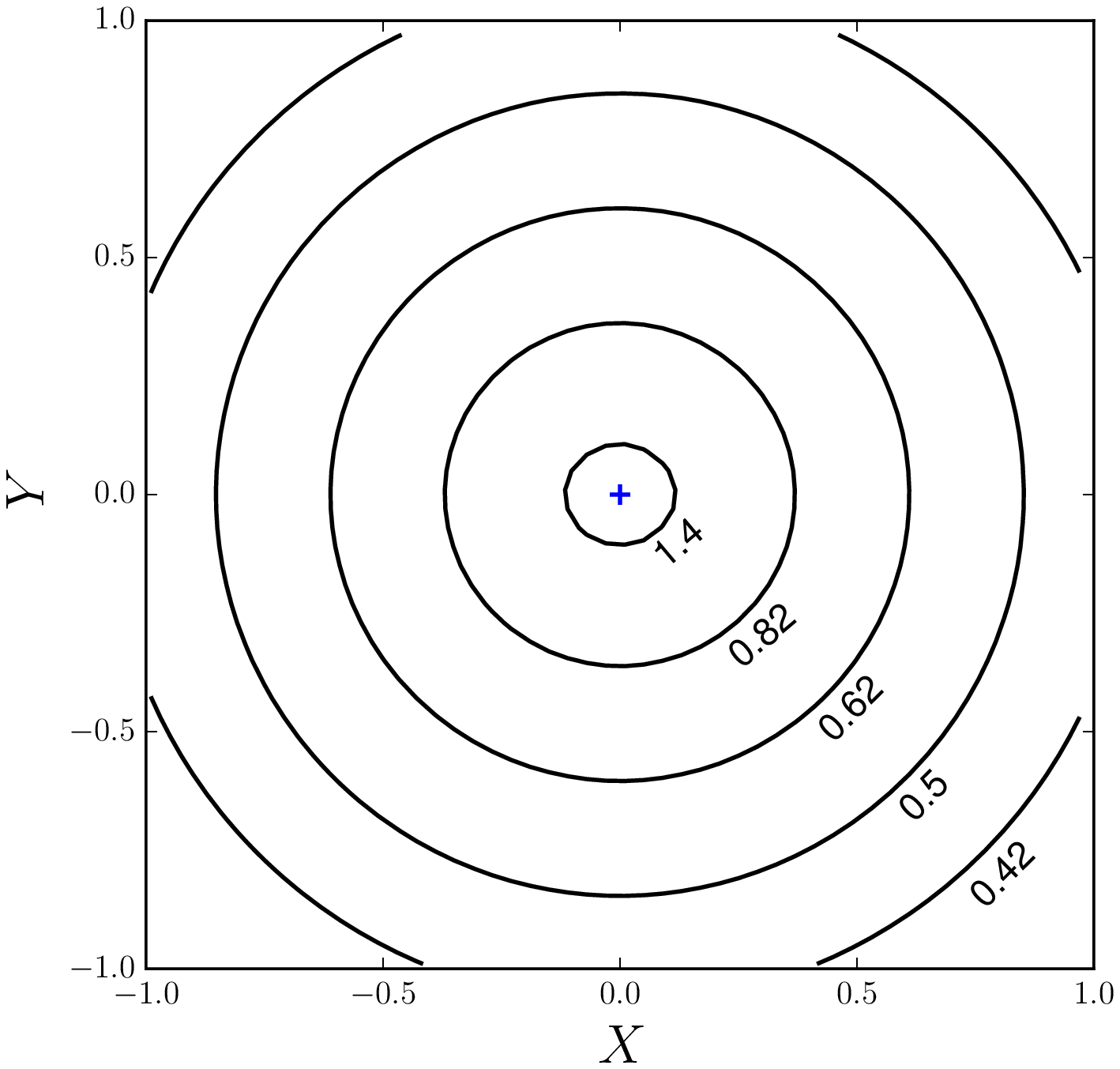}
\subcaption{ $i_{\rm o} = 45^{\circ}$ }
\end{subfigure}
\hfill
\begin{subfigure}{0.5\textwidth}
\centering
\includegraphics[width=0.9 \textwidth,trim={3cm 0cm 2cm 0cm}]{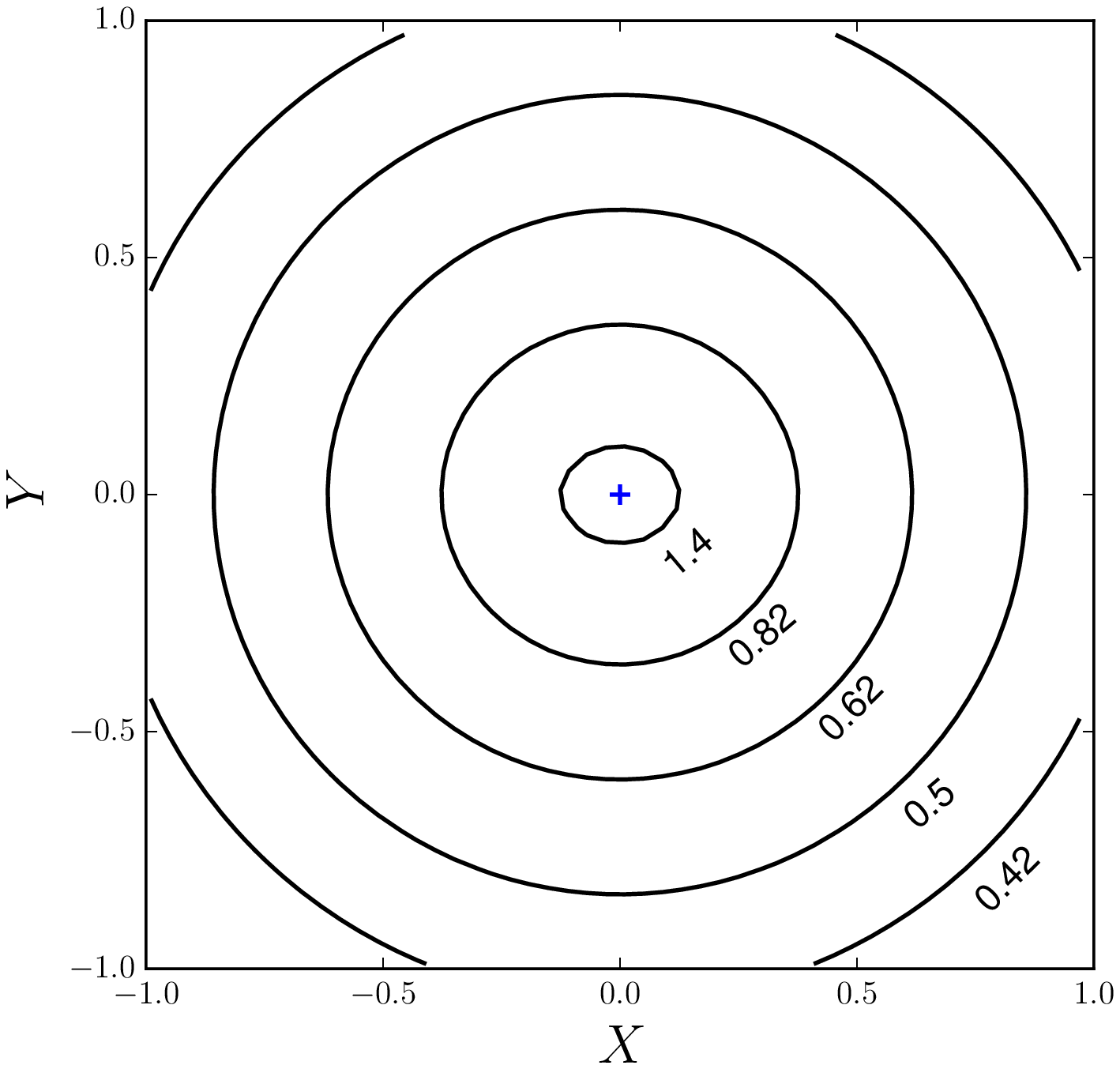}
\subcaption{ $i_{\rm o} = 90^{\circ}$ }
\end{subfigure}
\caption{ \emph{Surface density profile}, $\Sigma(X, Y)$ in units of 
$M_{\rm c}/\rc^2$, for two different viewing angles. Distances are measured in units of \rc.}
\label{sig-contours}
\end{figure}

\begin{figure}
\centering
\includegraphics[width = 0.6\textwidth]{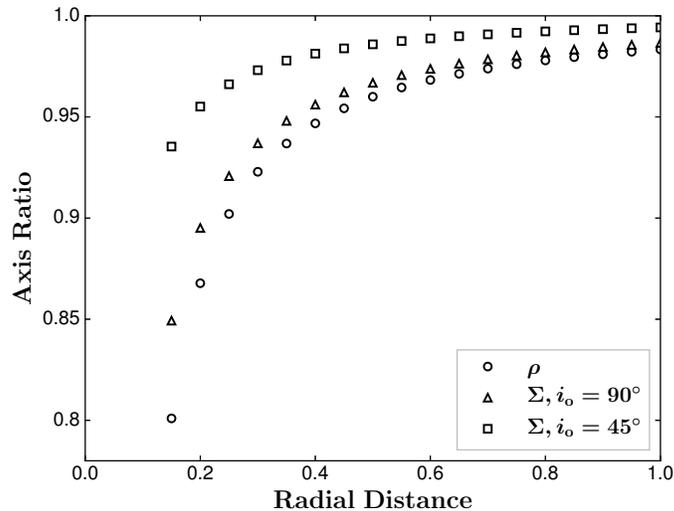}
\caption{Axis-ratio of the isocontours of total density, $\rho$, and 
surface density $\Sigma$, versus the major axis (in units of \rc) of the isocontours.}  
\label{axis-ratio}
\end{figure}

Figure~\ref{rho1} shows the isocontours of $\rho_1$ in the $(R,z)$ meridional plane, for $\gamma = 5/4$ and $n = -2\beta = 1/2$, for which ${\cal B}(1/2,5/4) = 2.41145$. The density perturbation $\propto r^{-5/2}$ rises steeply with decreasing $r$, similar to the density of the perturbing disc, $\rho_{\rm d}$. It is positive close to the equatorial plane of the disc (for $57.37^{\circ} < \theta < 122.63^{\circ}$) and negative otherwise, a property
that is independent of the cusp parameters $(n, \gamma)$. This behaviour is consistent with what we expected from the orbital dynamics discussed in the previous section. Figure~\ref{rho} plots the isocontours of the total density, $\rho(r,\theta) = \rho_{\rm c} + \rho_1$. These reveal an oblate spheroidal deformation of the spherical cusp. The flattening increases steeply with decreasing $r$, with the axis ratio $\sim 0.8$ at $\sim 0.15~\mbox{pc}$ --- see Figure~\ref{axis-ratio}. We also computed $\Sigma(X, Y)$, the surface density profile of the deformed cusp, by integrating $\rho(r,\theta)$ along different lines of sight upto a distance of $3~\mbox{pc}$ from the MBH, because this corresponds to the break-radius of the cusp \citep{gsd17}. Figure~\ref{sig-contours} shows the isocontours of $\Sigma$ on the sky plane for $i_{\rm o} = 45^{\circ}$ and $i_{\rm o} = 90^{\circ}$, where $i_{\rm o}$ is the angle between the line of sight and the disc normal.
The flattening increases steeply with decreasing $r$, similar to the density
profile; the edge-on view ($i_{\rm o} = 90^{\circ}$) shows maximal flattening, as can be seen from Figure~\ref{axis-ratio}.

\FloatBarrier

\section{Discussion and Conclusions}

We have presented a simple model of the deformation of a spherical stellar cusp (with anisotropic velocity dispersion) around a MBH, due to the growing gravity of a massive, axisymmetric accretion disc, for parameter values appropriate for the GC NSC. The mechanism is generic and may be common in galactic nuclei. 

We argued that the disc grows over times that are much longer than the typical apse precession period of cusp stars within a parsec of the MBH. The dynamical problem is not solvable in general stellar dynamics. But within $r_{\rm infl} \simeq 2\,\mbox{pc}$, the dominant gravitational force on a star is the Newtonian $1/r^2$ attraction of the MBH, and the semi-major axis of every star is an additional conserved quantity for evolution over several apse precession periods \citep{st99}. We used the secular theory of \citet{st16} to construct an integrable model of the adiabatic deformation of the cusp DF. Although the non-linear, self-consistent problem is integrable, the full solution requires a lot of numerical computations. In order to get an idea of the nature of the deformation, we used linear secular theory to obtain an analytical expression for the DF perturbation due to the `bare' effect of the disc. We explored orbital structure, which enables us to not only understand the physical properties of the linear deformation, but also to bound the limits of linear theory and discuss non-linear effects. The circulating orbits of linear theory are such that stars tend to spend more time near the equatorial plane of the disc, when their orbital eccentricity is maximal; this takes them closer to the inner, dense parts of the gas disc, an effect that could enhance the stripping of the envelopes of red giants \citep{ac14}.

Orbital structure also reveals the limits of linear theory, which does not apply to orbits whose apsides librate around $90^\circ$ or $270^\circ$. For any given $I$ and $L_z$, these orbits occupy regions of the highest eccentricities. Their DF depends on the entire orbital history --- in contrast to the orbits of linear theory which respect adiabatic invariance --- and requires computations based on the non-linear theory of adiabatic capture into resonance. For an initially tangentially anisotropic velocity dispersion, which seems to be the case for the GC NSC on scales $< 2~\mbox{pc}$ from the MBH \citep{feld17}, the relative number of eccentric orbits is small. Hence linear theory should do well as a first approximation for semi-major axes in the range $0.16 - 1~\mbox{pc}$.

{Secular stability is an important issue, which 
we now review in the light of earlier results for the linear dynamical stability of non-rotating spherical DFs, $F_0(I, L)$. For the lopsided $l=1$ 
linear mode \citet{tre05} showed that DFs with $(\partial F_0/\partial L) < 0$ are secularly stable, whereas DFs with $(\partial F_0/\partial L) > 0$
are either stable or neutrally stable when $F_0 =0$ at $L=0$ (i.e.
an empty loss-cone). The latter applies to the tangentially anisotropic case, $n = 1/2$, we have considered in this paper. \citet{pps07} considered 
mono-energetic DFs, $F_0(I, L) = \delta(I - I_0)f(L)$, dominated by nearly radial orbits. They found linear secular instabilities for $l\geq 3$ 
when $f(L)$ is a non--monotonic function of $L$. Relaxing the restriction
to nearly radial orbits, \citet{pps08} concluded that the non-monotonicity
of the DF as a function of $L$ is the main requirement for this (empty)
loss-cone instability to $l\geq 3$ modes. The cusp DFs of equation~(\ref{cusp-df}) are monotonic functions of $L$ for $n\neq 0$,
and may be expected to be stable in this sense; when $n=0$, the DF is 
a function only of $I$ and cannot be changed by any secular process because
$I$ is a secularly conserved quantity. So we are somewhat assured that the unperturbed cusp is likely to be linearly stable.} But this does 
not imply that an axisymmetric deformation, forced by a disc of small (but not infinitesimal) mass, is necessarily stable; it could runaway in an 
axisymmetric manner, or be vulnerable to the growth of non-axisymmetric 
modes. To investigate this aspect, we need to first include the effect of the self-gravity of the perturbation on its own evolution, and then 
explore the problem through $N$-body simulations.  

The density perturbation corresponding to the linear deformation results in an oblate spheroidal deformation of the formerly spherical cusp. The flattening increases steeply with decreasing distance from the MBH; the intrinsic axis ratio $\sim 0.8$ at $\sim 0.15~\mbox{pc}$. Surface density profiles for different viewing angles were presented. The appearance will
depend on the assumed plane of the gas disc, and one could consider this 
for the GC NSC. The planes of the young stellar disc close to the MBH,
and the CND farther away, have a high mutual inclination \citep{pgm06}. 
It is possible that the young stars were formed nearly coplanar with the 
CND and underwent dynamical evolution, also being perturbed by the CND
\citep{ssk09}. The ionizing radiation from the hot young stars also seems to have pushed gas out from beyond $0.5~\mbox{pc}$, and this would tend to decrease the spheroidal deformation we calculated at these distances. But a distinct possibility is that the accretion disc itself was warped. 

The gravitational perturbation of a warped gas disc would cause a non-axisymmetric deformation of the spherical cusp, so our calculation needs 
to be extended to account for this. We considered an unperturbed spherical stellar cusp with anisotropic velocity dispersion, because we wanted to 
begin the simplest generic case.\footnote{An isotropic secular DF, 
$F_0(I)$, cannot undergo any secular change, either through collisionless 
perturbations or through resonant relaxation, because $I$ is a secular 
invariant.} \citet{cfg15} constructed a self-consistent, flattened and rotating DF, $f(E, L_z)$, for the GC old stellar cusp. For 
$r < r_{\rm infl}$, this implies an unperturbed secular DF of the form, $F_0(I, L_z)$. Such a DF is immune to all secular axisymmetric perturbations, because $I$ and $L_z$ are conserved quantities for every stellar orbit. However, $F_0(I, L_z)$, would respond to the non-axisymmetric perturbation of a warped gas disc, because the $L_z$ of every orbit would then evolve with time, even though $I$ remains constant. The deformed cusp would then not be axisymmetric, a feature explored recently through triaxial modelling of the GC NSC \citep{feld17}.
    
\section*{Acknowledgments}
We would like to thank Tuan Do and Anna Ciurlo for helpful discussions, and the anonymous referee for insightful questions and comments.

\appendix
\section{Orbit-averaged disc potential}

In order to compute the orbit--averaged disc potential, 
$\Phi_{\rm d}(I, L, L_z, g, \tau)$, we need the following relations 
between $(r,z)$ and Keplerian orbital elements:
\beq
r \;=\; a \sqrt{1-e \, C_{\eta}}\,, \qquad \cos{\theta} \;=\; \frac{z}{r} 
\;= \; \frac{ S_{i}\left(S_{g} \, (C_{\eta} -e) \,+\, C_g \, \sqrt{1-e^2} \, S_{\eta} \right)}{1-e \, C_{\eta}}\,, 
\eeq
where $S$ and $C$ are shorthand for sine and cosine of the angle given 
as subscript, and $\eta$ is the eccentric anomaly. From equation~(\ref{disc-pot}), we see that the following three averages over the Kepler orbital phase, $w$, (or mean anomaly) need to be computed: 
$\left< \, 1/\sqrt{r} \, \right> $, $\left< \, \left|\cos{\theta}\right|/\sqrt{r} \, \right>$ and $\left< \, \cos^2{\theta}/\sqrt{r}  \, \right>$. 
Using $w = \eta - e  \sin{\eta}\,$ all of these can be expressed in 
terms of the elliptic integrals, listed below for ease of reference: 
\beq
{ \cal F}(\zeta_0,k) \;=\; \int_{0}^{\zeta_0} \rmd \zeta \, \frac{1}{\sqrt{1- k^2  \sin^2{\zeta}}}\,, \qquad\quad { \cal K}(k) \;=\; \int_{0}^{ \frac{\pi}{2}} \rmd \zeta \, \frac{1}{\sqrt{1- k^2  \sin^2{\zeta}}}\,,
\label{ell-int-1k}
\eeq
are incomplete and complete elliptic integrals of the first kind, and
\beq
{\cal E}(\zeta_0,k) \;=\; \int_{0}^{\zeta_0} \rmd \zeta \, \sqrt{1- k^2  \sin^2{\zeta}}\,, \qquad\quad {\cal E}(k) \;=\; \int_{0}^{ \frac{\pi}{2}} \rmd \zeta \, \sqrt{1- k^2  \sin^2{\zeta}}\,,
\label{ell-int-2k}
\eeq
are incomplete and complete elliptic integrals of the second kind.
Then the first average is:
\beq
\left< \frac{1}{\sqrt{r}}  \right> \;=\; \oint \frac{\rmd \eta}{2  \pi} \,\frac{(1-e \cos{\eta})}{\sqrt{r}} \;=\; \frac{1}{\pi  \sqrt{a}} \, \int_{0}^{\pi} \rmd \eta \, \sqrt{1-e  \cos{\eta}} \;=\; \frac{2  \sqrt{1+e}}{\pi  \sqrt{a}} \, {\cal E} \! \left( k \right)\,,
\label{avg-root-r}
\eeq
where $k(e) = \sqrt{2e/(1+e)}$. 

The second average is:
\beq
\left< \frac{ \left|\cos{\theta}\right|}{\sqrt{r}}\right> \;=\; \oint \frac{\rmd \eta}{2 \, \pi} \, (1 - e  \cos{\eta})  \frac{ | \cos{\theta} |}{\sqrt{r}} \;=\; \frac{\sin{i}}{\sqrt{a}} \int_{0}^{2 \pi} \frac{\rmd \eta}{2  \pi} \frac{ | S_g (C_{\eta}-e) +C_g  \sqrt{1-e^2}  S_{\eta} |  }{\sqrt{1- e \, C_{\eta}}}\,.
\label{avg-sroot-basic}
\eeq
Note that $| S_g (C_{\eta}-e) +C_g \, \sqrt{1-e^2} \, S_{\eta} |  = \sqrt{1- e^2  \cos^2{g}}  \;| \cos{(\eta- \eta_0) } - \cos{\theta_0} |$, where
\beq
\eta_0(e,g) \;=\; \tan^{-1}(\sqrt{1-e^2} \cot{ g}  )\,,\qquad\quad \theta_0(e,g) \;=\; \tan^{-1}\left( \frac{ \sqrt{1 - e^2}  }{e  
\vert\sin{g}\vert}  \right)\,.
\label{eta0-theta0-rels}
\eeq
In the angular interval $\eta \in [\eta_0 ,\, \eta_0 + 2\, \pi]\,$, the expression within ``$| \;|$'' changes sign at $\eta = \eta_0 + \theta_0$ 
and $\eta = 2 \pi + \eta_0 - \theta_0$. Rewriting  
\begin{align}
\left< \frac{ \left|\cos{\theta}\right|}{\sqrt{r}}\right> &\;=\; \frac{\sin{i}}{\sqrt{a}} \,\Bigg|\, \oint \frac{\rmd \eta}{2 \pi} \frac{ S_g  (C_{\eta} - e) \,+\, C_g  \sqrt{1-e^2}  S_{\eta}  }{\sqrt{1-e \, C_{\eta}}}
\nonumber\\[1 em]
&\qquad\qquad-\, 2  \int_{{\eta}_0 \,+\, {\theta}_0}^{2 \pi +{\eta}_0 -{\theta}_0} \frac{\rmd \eta}{2  \pi} \frac{ S_g  (C_{\eta} - e) \,+\, C_g  \sqrt{1-e^2}  S_{\eta}  }{\sqrt{1-e \, C_{\eta}}}\,\Bigg|\,, 
\end{align}
we obtain
\beq
\left< \frac{ \left|\cos{\theta}\right|}{\sqrt{r}}\right> \;=\; 
\frac{2  \sin{i}}{\pi  \sqrt{a}}  \,S(e,g)\,,
\label{avg-|c|-root-r} 
\eeq
where the function
\begin{align}
&S(e,g) \;=\; \frac{\sqrt{1+e}}{e}\,  |\sin{g}| \; \bigg[ - {\cal E}(k) \,+\,\, {\cal E}(\eta_2,k) \,-\, {\cal E}(\eta_1,k)\nonumber \\[1ex]
&\;\,+ (1-e)  \left\{ {\cal K}(k) - {\cal F}(\eta_2,k) + {\cal F}(\eta_1,k) \right\} \bigg] \;+\;\cos{g}  \frac{1-e^2}{e}  \bigg[ \frac{1}{ \sqrt{1-e \cos{g} }} - \frac{1}{\sqrt{1+e  \cos{g}}} \bigg]\,. 
\end{align}
Here $k$ is given below equation~(\ref{avg-root-r}), $\,(\eta_0 , \theta_0)$ are defined in equation~(\ref{eta0-theta0-rels}), and 
\beq
\eta_1 (e,g) \;=\; \frac{\eta_0(e,g) + \theta_0(e,g) - \pi}{2}\,, \qquad  \quad  \eta_2(e,g) \;=\; \frac{ \eta_0(e,g) - \theta_0(e,g) + \pi }{2}\,.   
\eeq
 
The last average is easier to do:
\begin{align}
\left< \frac{\cos^2{\theta}}{\sqrt{r}}  \right> &\;=\; \oint \frac{\rmd \eta}{2  \pi} \, (1-e  \cos{\eta})  \frac{\cos^2{\theta}}{\sqrt{r}} \;=\; \frac{\sin^2{i}}{\sqrt{a}}  \oint \frac{\rmd \eta}{2  \pi}  \frac{\left( S_g  (C_{\eta}-e) + C_g  \sqrt{1-e^2}  S_{\eta}  \right)^2}{(1-e  \, C_{\eta})^{\frac{3}{2}}}\nonumber\\[1 em]
& \;=\; \frac{2  \sin^2{i}}{\pi  \sqrt{a}} \left[ \frac{\sqrt{1+e}\, {\cal E}\!(k)}{2} \;-\; T(e)  \cos{2 g}  \right]
\label{avg-c2-root-r}
\end{align}

where the function
\beq
T(e) \;=\; \sqrt{1+e}  \left[ \left( \frac{2}{e^2}-\frac{3}{2}  \right) { \cal E}\!(k) - \frac{2}{e^2} (1-e) {\cal K}(k)  \right].
\eeq

\noindent
Using (\ref{avg-root-r}), (\ref{avg-|c|-root-r}) and (\ref{avg-c2-root-r}),
the orbit-averaged disc potential is:
\begin{align}
\Phi_{\rm d} &\;=\; \frac{16 G  \mdot }{11  \pi r_{\rm c}} \mu(\tau)\sqrt{\frac{r_{\rm d}}{a}}\, \Bigg[ -\frac{297}{100} \sqrt{1+e} \, {\cal E}\! \left( k \right)  \;+\; \frac{\sin{i}}{2}\,  S(e,g)
\nonumber\\[1em]
&\qquad -\, \frac{9}{100} \sin^2{i} \left( \frac{\sqrt{1+e}}{2}  {\cal E}\! \left(k \right) -  T(e) \cos{2g} \right) \bigg]. 
\label{avg-phi-d-exact}
\end{align}
This expression is used to compute the isocontours shown in Figure~1.
For dynamical calculations, we found it convenient to approximate the 
functions, $S(e,g)$ and $T(e)$, by the following polynomials in $e^2\,$:
\begin{align}
T(e) &\;\simeq\; a_t  e^2 + b_t  e^4+ c_t   e^6\,, \\
S(e,g) &\;\simeq \left(1 + a_0  e^2 + b_0  e^4 + c_0  e^6\right) \;-\;
\lambda\left(a_t  e^2 + b_t e^4 + c_t  e^6\right) \cos{2  g}\,,
\end{align}
where the constants,$(a_t,b_t,c_t,a_0,b_0,c_0,\lambda)$, are given below equation~(\ref{phid-avg}). This approximation results in a maximum error of 
$\sim 2 \%$ in $\Phi_{\rm d}$, and provides us with the simpler 
expression of equation~(\ref{phid-avg}).   

\section{Density deformation}

The density perturbation, $\rho_1 = M_{\rm c}/(2\pi)  \int F_1\, \rmd \bfu$, is defined by a triple-integral over velocities, of the DF 
perturbation, $F_1$, of equation~(\ref{f1-orb}). We use spherical
polar coordinates, with $\bfu = (u_r,u_{\theta},u_{\phi})$. The integrals can be transformed into integrals over $E$, $L$ and $L_z$ using the following relations:
\beq
L_z \;=\; r \sin{\theta}\,u_{\phi} \,,
\qquad  L \;=\; r \sqrt{u_{\theta}^2 + \frac{L_z^2}{ r^2  \sin^2{\theta} }}
\,,\qquad E \;=\; \frac{u_r^2}{2} + \frac{L^2}{2  r^2} - \frac{G \mdot}{r}\,.
\eeq
Then we have:
\beq
\rho_1(r,\theta) \;=\; \frac{2  M_{\rm c}}{\pi  r} \int_{-\frac{{GM}_\bullet}{r}}^{0} \rmd E  \int_{0}^{L_{\rm m}} \rmd L \frac{L}{ \sqrt{ L_{\rm m}^2 -L^2 }  } \int_{-L \sin{\theta}}^{L \sin{\theta}} \rmd L_z \frac{F_1}{ \sqrt{ L^2 \, \sin^2{\theta} - L_z^2  }  }\,,
\label{rho1-int}
\eeq
where $\,L_{\rm m }(E,r) = \sqrt{2  r^2  E + 2 G  \mdot  r}\,$ is the maximum
value of the (magnitude of the) angular momentum that an orbit of energy
$E$ can have at distance $r\,$. 

As $F_1 \propto \cos{2g}$, so we first express $\cos{g}$ in terms of $(\bfr, \bfu)$. Since $g$ is the angle between the ascending node and the periapse, we have: 
\beq
\cos{g} \;=\; \frac{1}{e \sqrt{L^2 - L_z^2}} \left[ \left( \frac{L^2}{G  \mdot} - r \right)(u_r \cos{\theta} \,-\, u_{\theta}\sin{\theta}) \,+\,r  u_r  \cos{\theta} \right]\,.
\eeq
Then
\beq
e^2  (L^2 - L_z^2)  \cos{2 g} = \mathscr{E}_1 + \mathscr{E}_2  ( L^2  \sin^2{\theta} - L_z^2) + \mbox{terms odd in $\bfu$}\,,
\eeq
where
\begin{align}
\mathscr{E}_1 &\;=\; L^2  \cos^2{\theta} \left[ \frac{2  L^2}{(G  \mdot)^2} \bigg(E - \frac{L^2}{r^2} + \frac{2G\mdot}{r} \bigg) \;-\; 1 \right]\,,
\\[1 ex]
\mathscr{E}_2 &\;=\; \frac{2}{r^2} \left( \frac{L^2}{G  \mdot} - r \right)^2 \;-\; e^2 \,.
\end{align}
Odd terms in $\bfu$ do not contribute to the $\bfu$-integral, so we can 
drop them. The integral over $L_z$ gives: 
\begin{align}
\mathscr{I}_1 &\;=\; \int_{-L\sin{\theta}}^{L \sin{\theta}} \rmd L_z  \frac{F_1}{ \sqrt{ L^2 \sin^2{\theta} - L_z^2  } }\nonumber\\[1em]
&\;=\; f_1  \bigg[ \frac{\lambda}{2  L} \int_{}^{} \rmd L_z  \frac{ \mathscr{E}_1 + \mathscr{E}_2  (L^2  \sin^2{\theta} - L_z^2)   }{ \sqrt{ (L^2  \sin^2{\theta} -L_z^2 ) (L^2-L_z^2) }}  - \frac{9}{100  L^2} \int_{}^{} \rmd L_z  \frac{ \mathscr{E}_1 + \mathscr{E}_2  (L^2  \sin^2{\theta} - L_z^2)    }{  \sqrt{L^2  \sin^2{\theta} -L_z^2} }\bigg]\,. \end{align}  
Although we have not shown it explicitly, the limits of the $L_z$-integrals
in the second line are the same as those in the first line. Here the factor, 
\beq
f_1 \;=\; \frac{2^{\frac{n}{2}} D(\tau)}{(G\mdot)^{n+\frac{1}{2}} \sqrt{\rc}}  \left(-E \right)^{n/2} L^{n-2}\left( a_t + b_t  e^2 + c_t e^4\right)\,. 
\eeq
The transformation, $ L_z = L  \sin{\theta}  \sin{\alpha}$, simplifies 
the integrals:
\begin{align}
\mathscr{I}_1 &\;=\; f_1  \bigg[ \frac{\lambda}{L^2}   \int_{0}^{\frac{\pi}{2}} \rmd \alpha \, \frac{\mathscr{E}_1 + \mathscr{E}_2  L^2  \sin^2{\theta}  \cos^2{\alpha}}{\sqrt{ 1- \sin^2{\theta}  \sin^2{\alpha}  }}    
\;-\; \frac{18}{100  L^2}  \int_{0}^{\frac{\pi}{2}} \rmd \alpha \, \left( \mathscr{E}_1 + \mathscr{E}_2  L^2  \sin^2{\theta}  \cos^2{\alpha}  \right) \bigg]\nonumber\\[1em]
&\;=\; f_1  \bigg[ \lambda  \left\{ \left( \frac{\mathscr{E}_1 }{L^2} - \mathscr{E}_2 \, \cos^2{\theta}  \right) {\cal K}(\sin{\theta}) + \mathscr{E}_2  {\cal E}(\sin{\theta})  \right\}
\;-\; \frac{9  \pi}{100 } \left( \frac{\mathscr{E}_1}{L^2} + \frac{ \mathscr{E}_2 \,  \sin^2{\theta}  }{2}  \right)
\bigg]\nonumber\\[1em]
&\;=\; 2  \pi  f_1 \bigg[ e^2 - \frac{2  L^2}{(G  \mdot  r)^2} (L_{\rm m}^2 - L^2)   \bigg]\Theta(\theta)\,, 
\end{align}
where 
\beq
\Theta(\theta) \;=\; \frac{\lambda}{2\pi}\left[\, {\cal E}(\sin{\theta}) \,-\, 2 \cos^2\theta \,{\cal K}(\sin{\theta})
\right] \;-\; \frac{9}{400}(1 \,-\,3\cos^2{\theta})\,. 
\eeq

The $L$-integral can be expressed in terms of Beta (B) functions:
\begin{align}
&\mathscr{I}_2 \;=\; \int_{0}^{L_{\rm m}} \rmd L \frac{L}{\sqrt{L_{\rm m}^2 -L^2}} \mathscr{I}_1\nonumber\\[1em]
&= \frac{2^{\frac{n}{2} + 1}  \pi  D(\tau)}{ (G  \mdot)^{n+\frac{1}{2}} \sqrt{\rc}     } \left(-E \right)^{\frac{n}{2}}  \Theta(\theta) \int_{0}^{L_{\rm m}} \rmd L \frac{L^{n-1}}{\sqrt{L_{\rm m}^2 - L^2}} (a_t + b_t  e^2 + c_t  e^4) \left[   e^2 - \frac{2  L^2 (L_{\rm m}^2 - L^2)}{(G  \mdot  r)^2} \right]\nonumber\\[1em]
&=  \frac{2^{\frac{n}{2} + 1}  \pi  D(\tau)}{ (G  \mdot)^{n+\frac{1}{2}} \sqrt{\rc}     } \left(-E \right)^{\frac{n}{2}}  \Theta(\theta) \bigg[  \frac{L_{\rm m}^{n-1}}{2} \bigg( \lambda_a  B_{\left( \frac{n}{2},\frac{1}{2} \right) } + (\lambda_b - \lambda_a) \frac{L_{\rm m}^2}{I^2} B_{\left(\frac{n}{2}+1,\frac{1}{2}\right)} + (\lambda_c - \lambda_b)  \frac{L_{\rm m}^4}{I^4} B_{\left(\frac{n}{2}+2,\frac{1}{2}\right)}
\nonumber\\[1em]
&\qquad\quad- \lambda_c  \frac{L_{\rm m}^6}{I^6} B_{\left(\frac{n}{2}+3,\frac{1}{2} \right)}  \bigg) 
  - \frac{L_m^{n+3}}{(G  \mdot  r)^2} \left( \lambda_a  B_{\left(\frac{n}{2}+1,\frac{3}{2} \right)  } + \lambda_b  \frac{L_{\rm m}^2}{I^2} B_{\left(\frac{n}{2}+2,\frac{3}{2} \right)} + \lambda_c \frac{L_{\rm m}^4}{I^4} B_{\left(\frac{n}{2}+3,\frac{3}{2} \right)} \right) \bigg]\,.
\end{align}

The final step is to the $E$-integral, $\rho_1 = (2  M_{\rm c}/\pi  r) \int_{-\frac{{GM}_\bullet}{r}}^{0} \rmd E \, \mathcal{I}_2\,$.
Substituting the explicit form for $L_{\rm m}$ given below the equation~(\ref{rho1-int}), and using $I = G  \mdot/ \sqrt{2(-E)}$, the integrals are once again given in terms of Beta functions. Therefore,
\beq
\rho_1(r,\theta,\tau) \;=\; \frac{3-\gamma}{4\pi}\,C_{n,\gamma}(\tau)\,\frac{M_{\rm c}}{\rc^3} \left(\frac{\rc}{r_{}}\right)^{\!\frac{5}{2}}\,\Theta(\theta)\,,  
\eeq
where
\begin{align}
C_{n,\gamma}(\tau) &\;=\; \frac{16n\,(2-\gamma)\, {\cal B}(n,\gamma)}{11\pi \, 2^{\left(\gamma-\frac{1}{2}\right)} \, \alpha_{\gamma}}\, \sqrt{\frac{r_{\rm d}}{\rc}}\,\mu(\tau)\,,\nonumber\\[1em]
{\cal B}(n,\gamma) &\;=\; 
 \frac{1}{ B_{\left(\frac{n}{2}+1,\frac{1}{2} \right)} \; B_{\left(\frac{2 \gamma +n-1}{2},\frac{n+3}{2}\right)}}     \Bigg[\lambda_a \, B_{\left(\frac{n}{2},\frac{1}{2} \right)} \; B_{\left( \frac{n}{2}+1,\frac{n+1}{2} \right)} \;+\; 2^2 (\lambda_b-\lambda_a) \, B_{\left( \frac{n}{2}+1,\frac{1}{2}\right)} \; B_{\left(\frac{n}{2}+2,\frac{n+3}{2}\right)}
\nonumber\\[1em]  
- 2^3  \lambda_a &\, B_{\left( \frac{n}{2}+1,\frac{3}{2} \right)} \; B_{\left( \frac{n}{2}+1, \frac{n+5}{2} \right) } \;+\; 2^4  (\lambda_c-\lambda_b) \, B_{ \left(\frac{n}{2}+2,\frac{1}{2} \right) } \; B_{\left(\frac{n}{2}+3,\frac{n+5}{2}\right)} \;-\; 2^5  \lambda_b \, B_{\left( \frac{n}{2}+2, \frac{3}{2} \right)} \; B_{\left( \frac{n}{2}+2,\frac{n+7}{2}\right) } 
\nonumber\\[1em] 
- 2^6  \lambda_c &\, B_{\left(\frac{n}{2}+3,\frac{1}{2}\right)} \; B_{ \left( \frac{n}{2}+4,\frac{n+7}{2}\right)} 
\;-\; 2^7  \lambda_c \, B_{\left( \frac{n}{2}+3,\frac{3}{2} \right) }\; B_{\left(\frac{n}{2}+3,\frac{n+9}{2} \right)} \Bigg]\,,\nonumber\\[1em]
\lambda_a &= a_t + b_t + c_t = 0.707106\,, \quad \lambda_b = -(b_t + 2 c_t) 
= -0.915737\,,\quad \lambda_c = c_t = 0.703998 \,.
\label{rho1-fin-app}
\end{align}

\end{document}